\documentclass[]{aa} %
\usepackage[]{natbib}
\usepackage{graphicx,color}
\usepackage{txfonts}
\usepackage[colorlinks=true, citecolor=blue]{hyperref}

\begin{document} 

\title{The enigmatic globular cluster UKS~1 obscured by the bulge: \textit{H}-band discovery of nitrogen-enhanced stars}

	\author{
		Jos\'e G. Fern\'andez-Trincado\inst{1,2}\thanks{Corresponding author: jose.fernandez@uda.cl},
		Dante Minniti\inst{3,4},
	    Timothy C. Beers\inst{5},
		Sandro Villanova\inst{6},
	    Doug Geisler\inst{6,7,8},
		Stefano O. Souza\inst{9},
		Leigh C. Smith\inst{10},
		Vinicius M. Placco\inst{11},
		Katherine Vieira\inst{1},
	    Angeles P\'erez-Villegas\inst{9},
	    Beatriz Barbuy\inst{9},
	    Alan Alves-Brito\inst{12},
        Christian Moni Bidin\inst{13},
        Javier Alonso-Garc\'ia\inst{14,15},
        Baitian Tang\inst{16}
        \and
        Tali Palma\inst{17,18}
		}
	
	\authorrunning{Jos\'e G. Fern\'andez-Trincado et al.} 
	
\institute{
         Instituto de Astronom\'ia y Ciencias Planetarias, Universidad de Atacama, Copayapu 485, Copiap\'o, Chile
         \and
         Institut Utinam, CNRS-UMR 6213, Universit\'e Bourgogne-Franche-Compt\'e, OSU THETA Franche-Compt\'e, Observatoire de Besan\c{c}on, BP 1615, 251010 Besan\c{c}on Cedex, France
         \and
         Depto. de Cs. F\'isicas, Facultad de Ciencias Exactas, Universidad Andr\'es Bello, Av. Fern\'andez Concha 700, Las Condes, Santiago, Chile
        \and
         Vatican Observatory, V00120 Vatican City State, Italy
        \and
         Department of Physics and JINA Center for the Evolution of the Elements, University of Notre Dame, Notre Dame, IN 46556, USA
         \and
         Departamento de Astronom\'\i a, Casilla 160-C, Universidad de Concepci\'on, Concepci\'on, Chile
         \and 
         Departamento de Astronom\'ia, Universidad de La Serena, Avenida Juan Cisternas 1200, La Serena, Chile
         \and
         Instituto de Investigaci\'on Multidisciplinario en Ciencia y Tecnolog\'ia, Universidad de La Serena. Benavente 980, La Serena, Chile
         \and
         Universidade de S\~ao Paulo, IAG, Rua do Mat\~ao 1226, Cidade Universit\'aria, S\~ao Paulo 05508-900, Brazil     
         \and 
         Institute of Astronomy, University of Cambridge, Madingley Road, Cambridge, CB3 0HA, UK
         \and
         NSF's Optical-Infrared Astronomy Research Laboratory, Tucson, AZ 85719, USA
         \and
         Universidade Federal do Rio Grande do Sul, Instituto de F\'isica, Av. Bento Gon\c{c}alves 9500, Porto Alegre, RS, Brazil
         \and
         Instituto de Astronom\'ia, Universidad Cat\'olica del Norte, Av. Angamos 0610, Antofagasta, Chile
         \and
         Centro de Astronom\'ia (CITEVA), Universidad de Antofagasta, Av. Angamos 601, Antofagasta, Chile
         \and
         Millennium Institute of Astrophysics, Santiago, Chile
        \and
        School of Physics and Astronomy, Sun Yat-sen University, Zhuhai 519082, China 
        \and
        Universidad Nacional de C\'ordoba, Observatorio Astron\'omico de C\'ordoba, Laprida 854, 5000 C\'ordoba, Argentina
        \and
        Consejo Nacional de Investigaciones Cient\'ificas y T\'ecnicas (CONICET), Godoy Cruz 2290, Ciudad Aut\'onoma de Buenos Aires, Argentina
   }
	
	\date{Received: 02/09/2020; Accepted: 22/09/2020}
	\titlerunning{Nitrogen-enhanced stars in UKS~1}
	
	
	\abstract
	{The presence of nitrogen-enriched stars in globular clusters provides key evidence for multiple stellar populations (MPs), as has been demonstrated with globular cluster spectroscopic data towards the bulge, disk, and halo. In this work, we employ the VVV Infrared Astrometric Catalogue (VIRAC) and the DR16 SDSS-IV release of the APOGEE survey to provide the first detailed spectroscopic study of the bulge globular cluster UKS~1. Based on these data, a sample of six selected cluster members was studied. We find the mean metallicity of UKS~1 to be [Fe/H]$=-0.98\pm0.11$, considerably more metal-poor than previously reported, and a negligible metallicity scatter, typical of that observed by APOGEE in other Galactic globular clusters. In addition, we find a mean radial velocity of $66.1\pm12.9$ km s$^{-1}$, which is in good agreement with literature values, within 1$\sigma$. By selecting stars in the VIRAC catalogue towards UKS~1, we also measure a mean proper motion of ($\mu_{\alpha}\cos(\delta)$, $\mu_{\delta}$) $=$ ($-2.77\pm0.23$,$-2.43\pm0.16$) mas yr$^{-1}$. We find strong evidence for the presence of MPs in UKS~1, since four out of the six giants analysed in this work have strong enrichment in nitrogen ([N/Fe]$\gtrsim+0.95$) accompanied by lower carbon abundances ([C/Fe]$\lesssim-0.2$). Overall, the light- (C, N), $\alpha$- (O, Mg, Si, Ca, Ti),  Fe-peak (Fe, Ni), Odd-Z (Al, K), and the \textit{s}-process (Ce, Nd, Yb) elemental abundances of our member candidates are consistent with those observed in globular clusters at similar metallicity. Furthermore, the overall star-to-star abundance scatter of elements exhibiting the multiple-population phenomenon in UKS~1 is typical of that found in other global clusters (GCs), and larger than the typical errors of some [X/Fe] abundances. Results from statistical isochrone fits in the VVV colour-magnitude diagrams indicate an age of 13.10$^{+0.93}_{-1.29}$ Gyr, suggesting that UKS~1 is a fossil relic in the Galactic bulge.
	}
	 
	\maketitle
	
	\section{Introduction}
	\label{section1}
	
	Globular clusters (GCs) are generally considered one of the key probes for revealing vital information about the mass-assembly history of the Milky Way \citep{Khoperskov2018, Minniti2018, Massari2019, Fernandez-Trincado2020, Hanke2020}. In the dawning era of the \textit{Gaia} mission \citep{Brown2018}, it has been possible to provide useful information on the fundamental parameters \citep[see, e.g.][]{Baumgardt2019} of almost all ($\sim159$) known GCs in the Milky Way \citep{Harris1999, Harris2010}. However, new observations in the near infrared (near-IR) from the VISTA Variables in the V\'ia L\'actea (VVV) ESO survey \citep{Minniti2010} have shown that the census of GCs in the central part of the Milky Way appears to be incomplete due to high interstellar extinction and crowding \citep{Alonso2017, Alonso2018}. A large number of GC candidates have been reported in the VVV survey \citep{Borissova2014, Minniti2017c}, and subsequent studies combining astrometric data from the \textit{Gaia} satellite \citep{Brown2018}, the VVV Infrared Astrometric Catalogue \citep[VIRAC:][]{Smith2018}, and chemistry from massive high-resolution spectroscopic surveys are helping to properly characterize these candidates \citep[see, e.g.][]{Contreras-Ramos2018, Villanova2019}. 
			
	In this context, near-IR high-resolution spectroscopy surveys such as the Apache Point Observatory Galactic Evolution Experiment \citep[APOGEE][]{Majewski2017} have helped minimize the effect of extinction, allowing for the detailed study of the Galactic bulge region \citep[see also][]{Rojas-Arriagada2020, Queiroz2020a}. Detailed abundances for a number of chemical species with a variety of nucleosynthetic origins have provided deeper insight into the multiple-population phenomenon \cite{Carretta2003, Bastian2018, Szabolcs2020} found in virtually all bulge GCs \citep{Carretta2009a, Pancino2017, Szabolcs2020}. Identifying and/or testing the mechanism responsible for this puzzling phenomenon in GCs \citep[see, e.g.][]{Decressin2007, Mink2009, Bastian2018} helps us understand not only GC formation and evolution, but also the chemical evolution of galaxies \citep{Pipino2009, Mauro2013, Barbuy2018}.
	
	 It is now firmly established that stars showing light-element abundance variations are almost ubiquitous within GCs \citep[see, e.g.][]{Gratton2004, Carretta2009a, Szabolcs2020}. The most commonly measured signatures are the Na-O anti-correlation \citep[see, e.g.][and references therein]{Carretta2009a}, the Al-Mg anti-correlation \citep{Pancino2017}, and the N-C anti-correlation \citep{Masseron2019, Szabolcs2020}. Stars within a cluster often exhibit higher [Na,N/Fe] and lower [C,O/Fe] values, with the former well above the typical Galactic levels at the same [Fe/H] as the cluster. In this context, \citet{Schiavon2017b} and \citet{Fernandez-Trincado2019d} have demonstrated that the identification of stars in bulge GCs with stellar atmospheres strongly enriched in nitrogen is a reliable tracer of the multiple-population phenomenon at all metallicities, and that it prevails in bulge GCs as metal-rich as [Fe/H]$\sim-0.1$ \citep[see also][]{Cohen1999, Tang2017}.	 
	 
	 	\begin{figure}	
	 	\begin{center}
	 		\includegraphics[width=90mm]{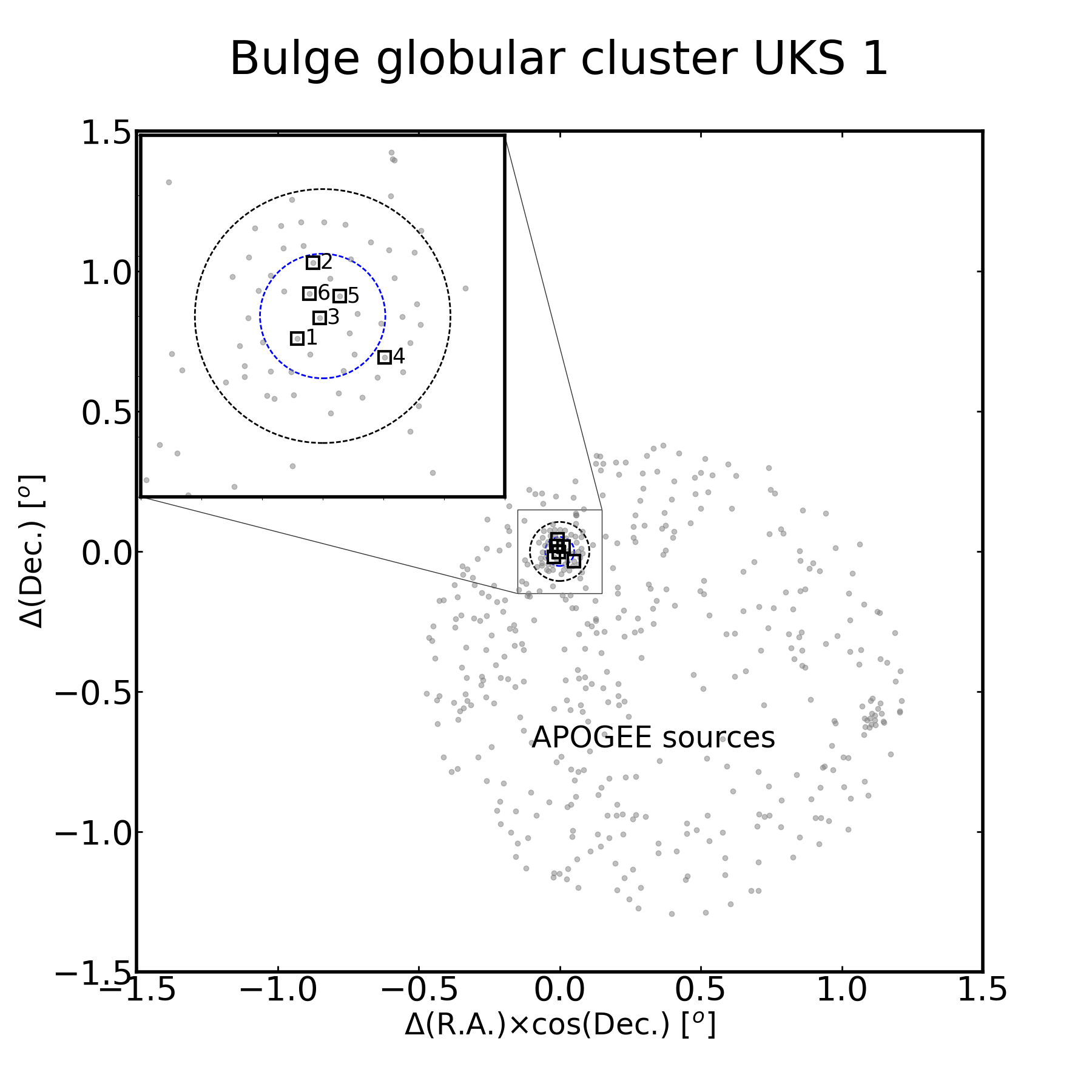}
	 		\caption{Spatial distribution of stars (grey dots) in the APOGEE-2 survey towards the bulge GC UKS~1. The highest likelihood members of UKS~1 are marked with black open squares. The black circle is the tidal radius of UKS~1 ($\sim$6.32') at $\sim$7.8 kpc \citep{Baumgardt2019}; the blue dotted circle indicates the tidal radius ($\sim$3.1') at $\sim$15.9 kpc \citep{Minniti2011}.}
	 		\label{Figure1a}
	 	\end{center}
	 \end{figure}	
	 
	 Despite the enormous observational efforts being carried out towards the bulge region, e.g. CAPOS (the bulge Cluster APOgee Survey -- Geisler et al. 2020, in preparation), a number of GCs have not been thoroughly investigated due to high foreground extinction. This also strongly limits observations in the optical regime \citep[see, e.g.][]{Cohen2018c}, pending independent constraints from ground-based spectroscopic observations. Among them is the case of UKS~1, a bulge GC discovered by \citet{Malkan1980} and neighbour of VVV CL001 \citep{Minniti2011}, which lies in a region of the Galactic bulge where interstellar extinction is very high, with E(B-V) $\sim$2.2 -- 3.10 \citep{Minniti1995, Bica1998, Harris1999, Ortolani2001, Minniti2011}. Only the red giant branch (RGB) has been detected so far in near-IR colour-magnitude diagrams (CMDs)\citep{Minniti1995, Ortolani1997}; furthermore, the fainter horizontal branch (HB) and the sub-giant branch (SGB) region were just barely reached with HST NICMOS photometry \citep{Ortolani2001}. This implies that UKS~1 is a very distant cluster ($\sim$9.3 -- 15.9 kpc), possibly located beyond the Galactic centre \citep[see, e.g.][]{Ortolani2007, Minniti2011}. More recently, \citet{Baumgardt2019} estimated a smaller heliocentric distance of $\sim$7.8 kpc and a cluster mass of $\sim0.8\times10^{5}$ M$_{\odot}$ by comparing the cluster density profile from the most recent \textit{Gaia} DR2 data \citep{Brown2018} to a large suite of direct N-body star cluster simulations. There is also a discrepancy regarding the cluster metallicity, that is to say, integrated infrared photometry indicates metallicities of [Fe/H] $\sim-1.2$ and $-1.18 $ \citep{Bica1998}, while echelle spectra covering the range 1.5 -- 1.8 $\mu$m estimate an intermediate metallicity of [Fe/H]$=-0.78$ \citep{Origlia2005}. In addition, there is no age estimate available for UKS~1. Therefore, a detailed and reliable physical characterization of this cluster is still lacking \citep{Cohen2018c}.
	 
	 The main purpose of the present paper is to explore the near-IR high-resolution spectroscopic observations from the APOGEE survey towards UKS~1, combined with an updated version of the VVV Infrared Astrometric Catalogue (VIRAC). These help us obtain a better view of the properties of the stellar population in the inner regions of UKS~1, as well as to produce a much improved CMD, helping to roughly estimate the cluster age by adopting a Bayesian statistical approach \citep[see, e.g.][]{Souza2020}. In Section \ref{section2}, we briefly describe the observations. In Section \ref{section3}, the data and selection of the potential members of UKS~1 are described. In Section \ref{section4}, we describe the adopted atmospheric parameters, followed by a discussion of the chemical abundances of our candidate stars in Section \ref{section5}. We present an orbital analysis of UKS~1 in Section \ref{section6}. In Section \ref{section10} we present results for the estimated age of UKS~1 based in a statistical isochrone fitting method following a Bayesian approach. Finally, in Section \ref{section7}, we summarize the results and draw our conclusions. The proper motion computation from VIRAC data and the differential reddening correction are described in Appendices \ref{section8} and \ref{section9}.
	 
	\section{UKS~1 in the APOGEE DR16}
	\label{section2}

	UKS~1 was observed by one of the APOGEE twin spectrographs \citep{Wilson2012, Eisenstein2011, Wilson2019} mounted on the Ir\'en\'ee du Pont 2.5m telescope at Las Campanas Observatory \citep{Bowen1973} in Chile. These observations were part of the APOGEE bulge programme with field centre in ($l$ , $b$) $\sim$ (5$^{\circ}$, 0$^{\circ}$), collecting high-resolution ($R\sim$22,500) \textit{H}-band spectral information for 448 sources. Cluster targets were positioned towards the north-west direction of the field in (~$l$,~$b$)~$\sim$~(~5.13$^{\circ}$,~0.76$^{\circ}$), and four visits were needed in order to archive a minimal S/N $>50$ at K$_{\rm s, 2MASS} \lesssim 12.5$ mag \citep[see][for details regarding the sample selection]{Zasowski2017}. The spectra of the potential cluster stars analysed in this work have S/N between 96 to 192. This work makes use of public spectra collected in the sixteenth data\footnote{For further details, we direct the reader to \citet[][]{GarciaPerez2016a} -- APOGEE Stellar Parameter and Chemical Abundances pipeline (ASPCAP), \citet{Holtzman2018} -- grid of synthetic spectra and discussion of associated errors, and \citet{Nidever2015} -- data reduction pipeline for APOGEE. The model grids for APOGEE DR16 are based on a complete set of \texttt{MARCS} \citep{Gustafsson2008} stellar atmospheres, which are now extended T$_{\rm eff}\lesssim$ 3200K.} release \citep[APOGEE DR16:][]{Ahumada2020}, as part of the Sloan Digital Sky Survey IV \citep[][]{Blanton2017}. 
	
	\section{Membership}
    \label{section3}
    	
    We identified a sample of potential stellar members of UKS~1 in the APOGEE DR16 database. Table \ref{Table1} summarizes the main properties of the likely members of UKS~1. Radial velocities are taken from the APOGEE catalogue \citep{Ahumada2020}, while the absolute proper motions and near-IR magnitudes (J, K$_{\rm s}$) are retrieved from VIRAC and 2MASS, respectively.
       
    We selected probable cluster members based on their radial velocities, absolute proper motions, metallicity, and position in the near-IR CMD. First, we restrict our sample to stars within 6.32$'$ from the GC centre, that is, inside the nominal GC tidal radius given in the new catalogue of fundamental properties of Galactic GCs \citep{Baumgardt2018, Baumgardt2019, Hilker2020}. 
    
    In addition, for the stars selected above, we manually validated their metallicities ([Fe/H]) from the strength of selected iron (Fe I) lines, by adopting the same methodology as described in \citet{Fernandez-Trincado2019b}, i.e. the [Fe/H] ratios have been derived by using the \texttt{BACCHUS} code \citep{Masseron2016}, and by performing a LTE analysis with a MARCS grid of spherical models \citep{Gustafsson2008}. For the remainder of this work, metallicity refers to the [Fe/H] determined from Fe I absorption features. We refer the reader to \citet{Nataf2019} and \citet{Szabolcs2020} for further discussion of differences between metallicities from optical and IR range.
   
\begin{figure*}	
     	\begin{center}
     		\includegraphics[width=190mm]{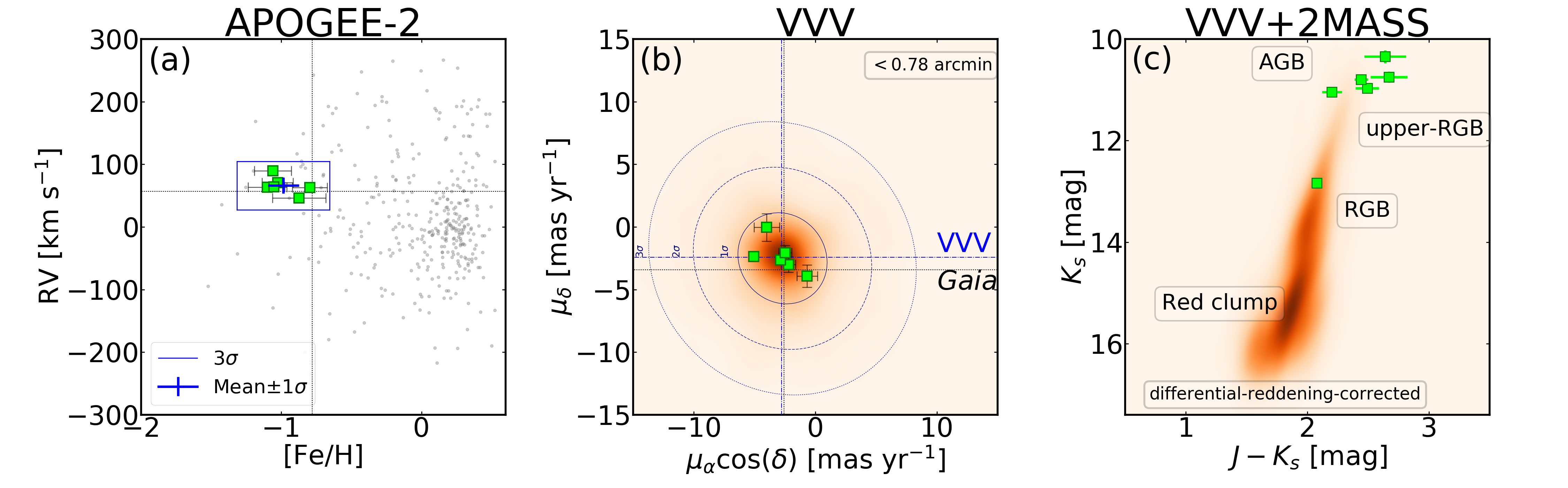}
     		\caption{(a) Radial velocity versus [Fe/H] for APOGEE stars towards the UKS~1 field. (b) kernel density estimation (KDE) of the VIRAC proper motions. (c) CMD using VVV$+$2MASS J and K$_{s}$ bands for stars within the half-light cluster radius. The six highest likelihood members of UKS~1 analysed in this work are marked with lime square symbols. In (a), the average in radial velocity and metallicity of our sample is marked by the blue cross symbol, while the blue rectangle indicates 3$\sigma_{\rm RV}$ and 3$\sigma_{\rm [Fe/H]}$. The black dotted lines in (a) and (b) indicate the nominal [Fe/H], $\mu_{\alpha}\cos{\delta}$, $\mu_{\delta}$, and radial velocity from \citet{Origlia2005} and \citet{Baumgardt2019}, respectively. The blue dotted lines in (b) indicate our estimated $\mu_{\alpha}\cos{\delta}$ and $\mu_{\delta}$ using VIRAC proper motions. Error bars are provided in Tables \ref{Table1} and \ref{Table2}.}
     		\label{Figure2}
     	\end{center}
     \end{figure*}

     Then, we selected objects that clump in the [Fe/H]--radial velocity and proper motion (PM) space, located closest to the nominal metallicity, radial velocity, and PMs of the cluster. The final sample contained six potential cluster members. The sky position of these stars is displayed in Figure \ref{Figure1a}; their proper motion distribution, metallicity, and radial velocities are summarized in Figure \ref{Figure2}. 
     
     		\begin{figure*}	
     	\begin{center}
     		\includegraphics[width=170mm]{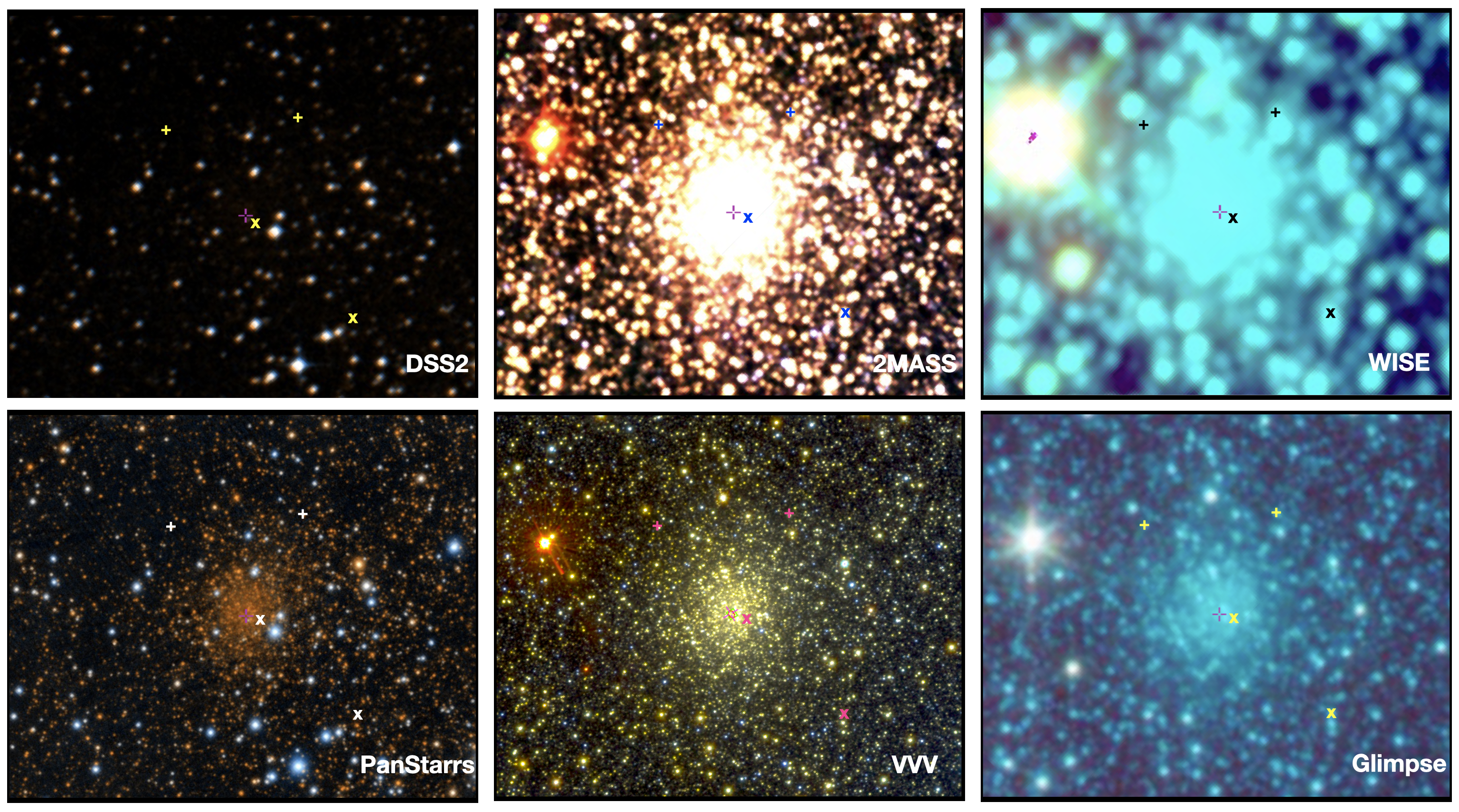}
     		\caption{From left to right, and from top to bottom: DSS2, 2MASS, WISE, PanStarrs, VVV, and Glimpse images (size 6'$\times$5') of the region containing the GC UKS~1. The `X' (N-rich stars) and `+'  (N-normal stars) signs in each panel mark the position of 4 out of the 6 highest likelihood members of UKS~1 analysed so far in this work; the magenta plus symbol indicate the centre of UKS~1.}
     		\label{Figure1b}
     	\end{center}
     \end{figure*}	
     
     Figure \ref{Figure1b} follows the sky position of 4 out of the 6 stars in a mosaic of multi-band images (FoV 6'$\times$5') of six surveys, including DSS-2, 2MASS, WISE, Pan-Starrs, VVV, and Glimpse. UKS~1 is not detected in DSS-2 due to high extinction. It shows up in full detail in the near-IR with VVV and Glimpse. It becomes a conspicuous cluster in Pan-Starrs, but is very contaminated in 2MASS, Glimpse, and WISE. We note that one of our stars identified so far in the very innermost region of the cluster has an atmosphere strongly enriched in nitrogen, supporting the existence of multiple stellar populations (MPs) in UKS~1 (as summarized below ). 
      
     Finally, we examined the location of our six potential cluster members in the differential-reddening corrected VVV$+$2MASS CMD. Figure \ref{Figure2} plots the CMD of the inner region (within 0.78' of the cluster centre) from the VVV$+$2MASS data, which clearly shows the RGB morphology of UKS~1. The UKS~1 RGB stars do not appear to exhibit sufficient scatter to accommodate a wide range in [Fe/H], in agreement with other GCs at similar metallicity. We find that the selected stars from the APOGEE catalogue lie in the upper part of the RGB, indicated by lime squares in Figure \ref{Figure2}. 
    
      \section{Spectroscopic parameters}
    	\label{section4}
 	
       To determine the chemical abundances of the selected members from UKS 1, we adopted the uncalibrated stellar atmospheric parameters (T$_{\rm eff}$, $\log$ \textit{g}, and [M/H]) computed from the \texttt{ASPCAP} pipeline \citep{GarciaPerez2016a}. With the fixed T$_{\rm eff}$, $\log$ \textit{g}, and first-guess [M/H], the first step consisted in determining the metallicity from selected Fe I lines, the micro-turbulence velocity ($\xi$), and the convolution parameter with the \texttt{BACCHUS} code \citep{Masseron2016}. Thus, the metallicity provided is the average abundance of selected Fe lines, while the micro-turbulence velocity is obtained by minimizing the trend of Fe abundances against their reduced equivalent width, and the convolution parameter stands for the total effect of the instrument resolution. We proceed in the same manner as in \citet[][]{Hawkins2016}, i.e. we derive a single global convolution value per spectrum, based on the average broadening of Fe lines, by assuming a Gaussian convolution profile. Once [Fe/H], $\xi$, and the convolution parameters are determined, O, C, and N abundances are estimated from selected $^{16}$OH, $^{12}$C$^{16}$O, and $^{12}$C$^{14}$N molecules \citep{Smith2013}. Once those elements are measured, the full process is iterated until convergence \citep[see, e.g.][]{Fernandez-Trincado2016, Fernandez-Trincado2017, Fernandez-Trincado2019a, Fernandez-Trincado2019b, Fernandez-Trincado2019c, Fernandez-Trincado2019d}. 
    
       Finally, for each chemical species and each line, the abundance ratios are determined with the \texttt{BACCHUS} code following the procedure as described in \citet[][]{Hawkins2016}, and briefly summarized here for guidance. (\textit{i}) a spectrum synthesis, using the full set of lines from the internal APOGEE DR14 atomic/molecular linelist (linelist 20150714) to find the local continuum level via a linear fit;  (\textit{ii}) cosmic and telluric rejections are performed;  (\textit{iii}) the local S/N is estimated;  (\textit{iv})  a series of flux points contributing to a given absorption line is automatically selected; and (\textit{v}) abundances are then derived by comparing the observed spectrum with a set of convolved synthetic spectra characterized by different abundances. The code finally proceeds with four different abundance determinations: (\textit{a}) line-profile fitting; (\textit{b}) core line intensity comparison; (\textit{c}) global goodness-of-fit estimate ($\chi^{2}$); and (\textit{d}) equivalent-width comparison. Each method yields validation flags, and a decision tree then rejects the line or accepts it, keeping the best-fit abundance. In this work, we adopt the $\chi^{2}$ method, which is most robust \citep[e.g.][]{Hawkins2016, Fernandez-Trincado2019b}. However, the information from the other diagnostics is stored, including the standard deviation between all four methods.
       
       Using the methods outlined above, we re-analyse the APOGEE spectra of the six cluster member candidates and manually estimate their chemical abundances. In this work, we focus on the Fe-peak elements (Fe, Ni), the light-elements (C, N), the $\alpha-$elements (O, Mg, Si, Ca, and Ti), the odd-Z elements (Al, K), and the \textit{s}-process elements (Ce, Nd, Yb) because their spectral lines are relatively strong and prominent in the APOGEE spectra for mildly metal-poor stars. Table \ref{Table1} shows the derived chemical abundances. The reference Solar photospheric abundances are from \citet{Asplund2005}, except for Ce II, Nd II, and Yb II, for which we have adopted the Solar abundances from \citet{Grevesse2015}. The adopted atmospheric parameters of our stars are consistent within the error range with parameters of UKS~1 RGB stars analysed in \citet{Origlia2005}.
       
       It is important to note that we do not provide abundance determinations based on photometric temperatures, as this introduces its own set of problems, mostly related to high E(B-V) values \citep[see, e.g.][]{Szabolcs2020}, and particularly because UKS~1 lies in a region with significant reddening (E(B$-$V)$\gtrsim$ 2.2). For this reason, we limited our analysis to uncalibrated stellar parameters from \texttt{ASPCAP}, as any photometric temperature estimation will not be reliable towards UKS~1. 
      
        In order to further check the statistical significance of our method in the calculation of chemical abundances, we also compute synthetic spectra by assuming uncertainties in the parameters in the range of $\Delta$T$_{\rm eff} = 100$ K, $\Delta \log $ \textit{g}$=0.3$, and $\Delta\xi = 0.05$ km s$^{-1}$, following the strategy outlined in \citet{Fernandez-Trincado2019b}, to investigate the sensitivity of abundances due to the variations in the adopted atmospheric parameters. Table \ref{Table2} shows the mean sensitivity of abundances according to the atmospheric parameters changes. The variation of atmospheric parameters result in a total uncertainty of about 0.1 to 0.25 dex in abundances; the effective temperature and $\log$ \textit{g} uncertainties, along with line-by-line variation, are the main contribution of the abundance uncertainties, depending on the chemical species.
       
\begin{figure*}	
        	\begin{center}
        		\includegraphics[width=125mm]{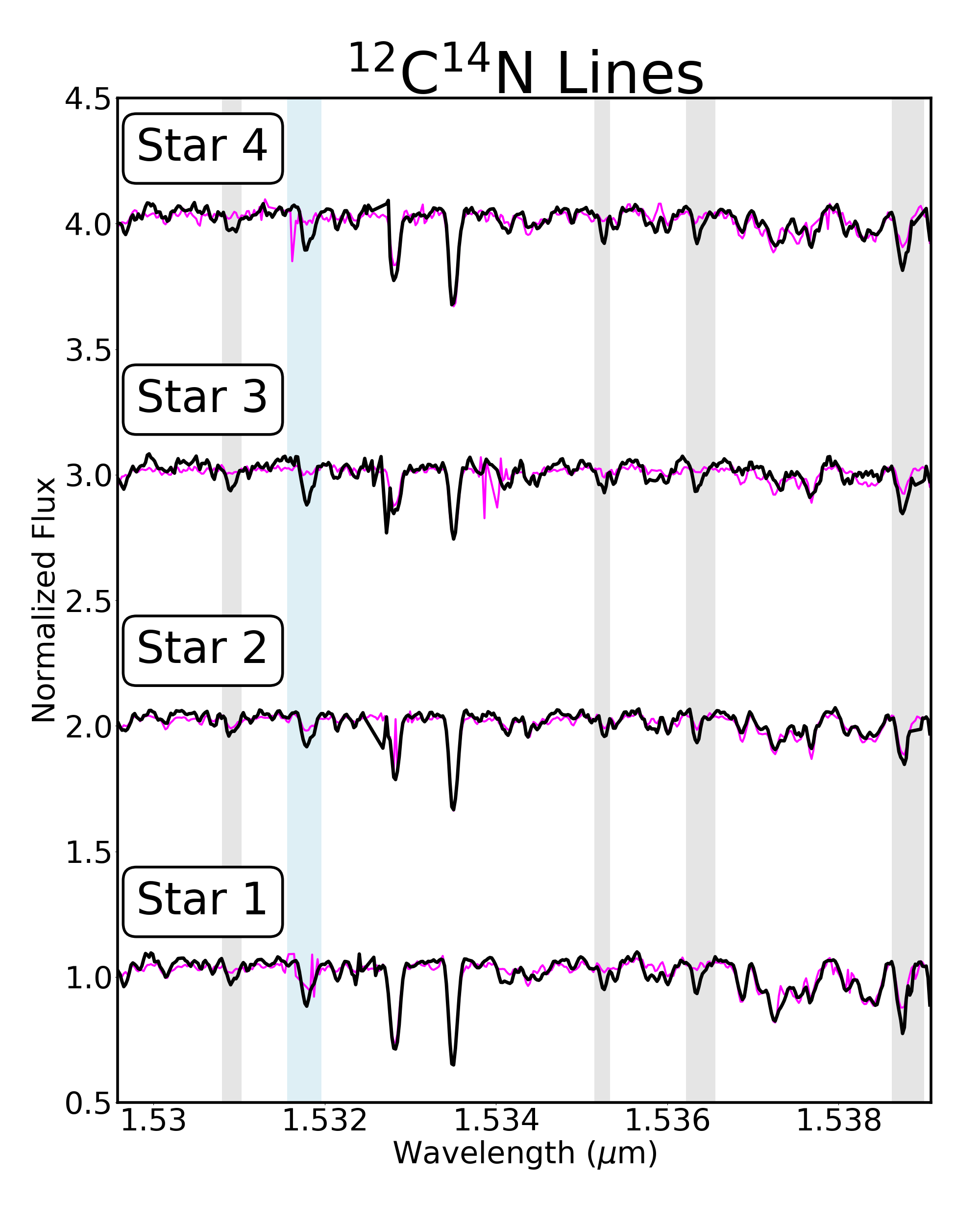}\includegraphics[width=49mm]{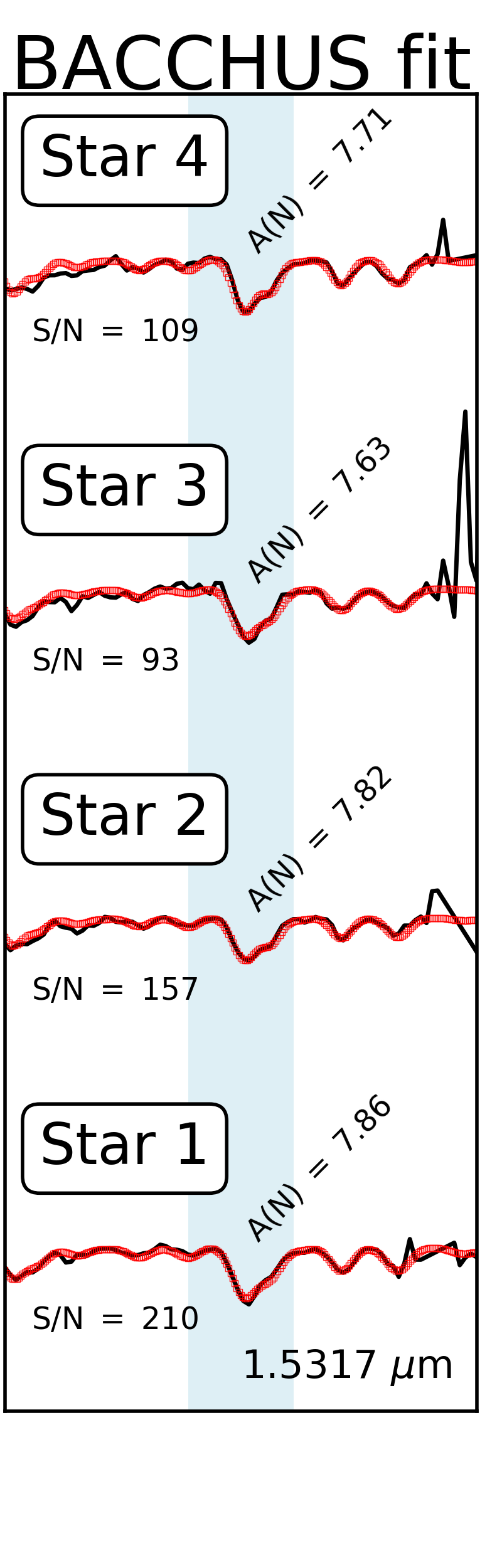}
        		\caption{\textit{Left:} APOGEE spectra (in air wavelength) in a region containing several $^{12}$C$^{14}$N lines, indicated by vertical shaded regions. The magenta line show the spectrum of N-normal stars with similar atmospheric parameters as the N-rich stars. \textit{Right:} Example of our spectral synthesis analysis (red squares) around the line 1.5317 $\mu$m (blue light shadow region). The A(N) at 1.5317 $\mu m$ is marked.}
        		\label{Figure3}
        	\end{center}
        \end{figure*}

       Based on the calculated chemical abundances, we have discovered that four of our six likely cluster members are nitrogen-enriched. Figure \ref{Figure3} provides a brief examination of typical \textit{H}-band spectra for the four nitrogen enriched stars around the $^{12}$C$^{14}$N spectral absorption features. This figure confirms the existence of a real chemical peculiarity in these objects. The spectra of these stars are shown in a wavelength range containing several $^{12}$C$^{14}$N lines, which are indicated by the grey and cyan shadow regions. The N-rich stars have remarkably stronger $^{12}$C$^{14}$N lines compared to a star with similar atmospheric parameters and a normal ([N/Fe]$\lesssim+0.5$) nitrogen abundance, which, in view of the fact that the pair of stars have nearly the same atmospheric parameters, can only mean that the UKS~1 stars must have much higher nitrogen abundances. The same figure shows an example of the corresponding $^{12}$C$^{14}$N lines modelled with the \texttt{BACCHUS} code.
        
      \section{Chemical properties of UKS~1}
       \label{section5}
 
      The present study adds a substantial contribution to the chemical characterization of UKS~1. In comparison, \citet{Origlia2005} analysed the elemental abundances [Fe/H], [O/Fe], [Ca/Fe], [Si/Fe], [Mg/Fe], [Ti/Fe], [$\alpha$/Fe], and [C/Fe] of four likely cluster members at a similar (1.5 - 1.8 $\mu m$) spectral coverage. Our sample of likely cluster members increases this number to six, and to the best of our knowledge, this is the largest sample yet analysed in UKS~1, allowing us to observe the abundance variations comparatively within the cluster. Table \ref{Table1} contains the abundance values of the fourteen chemical species ([C/Fe], [N/Fe], [O/Fe], [Mg/Fe], [Al/Fe], [Si/Fe], [K/Fe], [Ca/Fe], [Ti/Fe], [Fe/H], [Ni/Fe], [Ce/Fe], [Nd/Fe], and [Yb/Fe]) determined in this work. 
      
      We proceed to compare our results with the homogeneous sample from \citet{Szabolcs2020}, as that data set has been carefully examined with the same code and similar methodology as adopted in this paper. We avoid any comparison with GCs only based in the \texttt{ASPCAP} APOGEE pipeline as they may exhibit larger systematic offsets \citep[see, e.g.][]{Nataf2019}. 
       
       \begin{figure*}	
     	\begin{center}
     		\includegraphics[width=95mm]{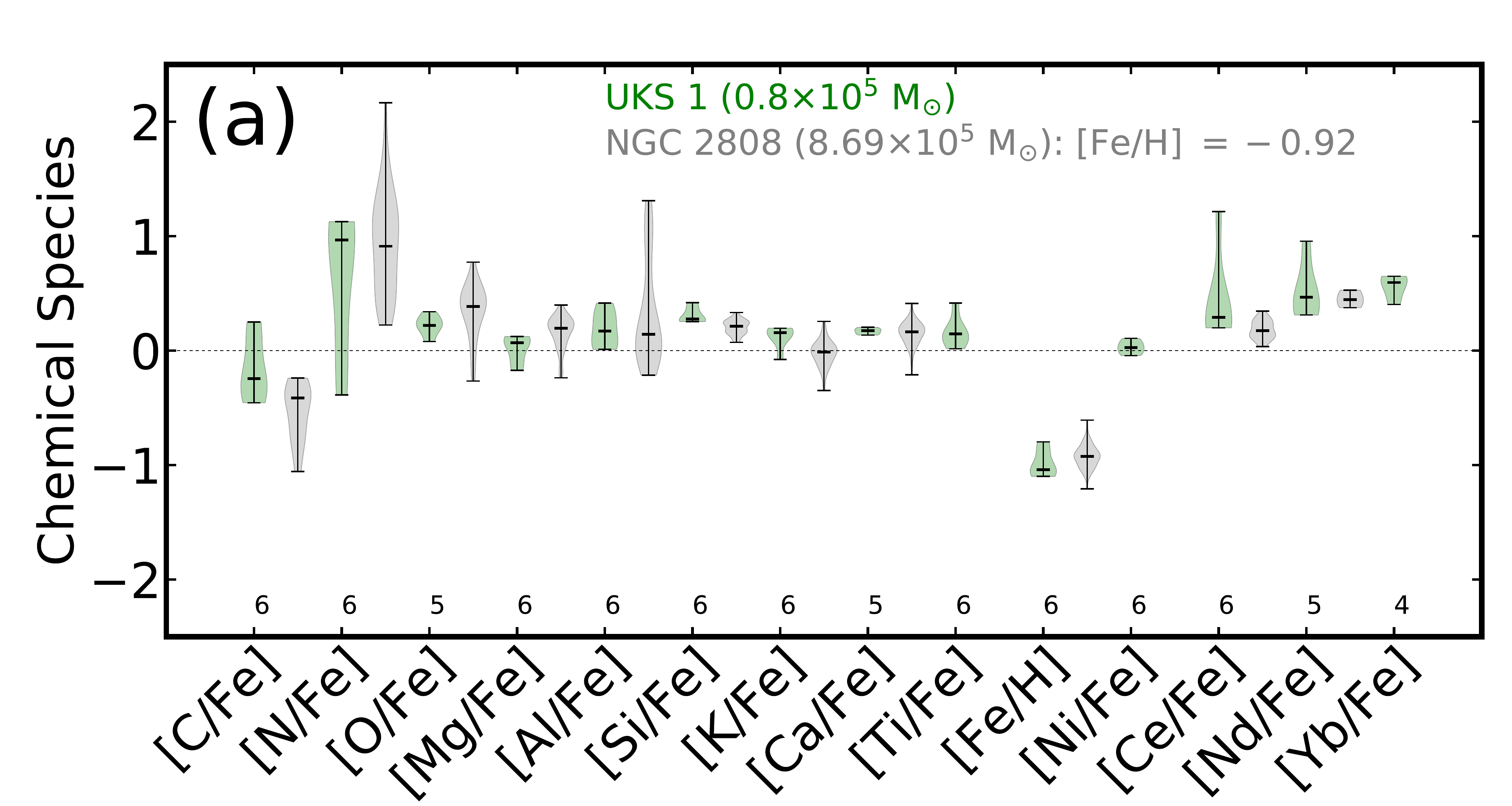}\includegraphics[width=95mm]{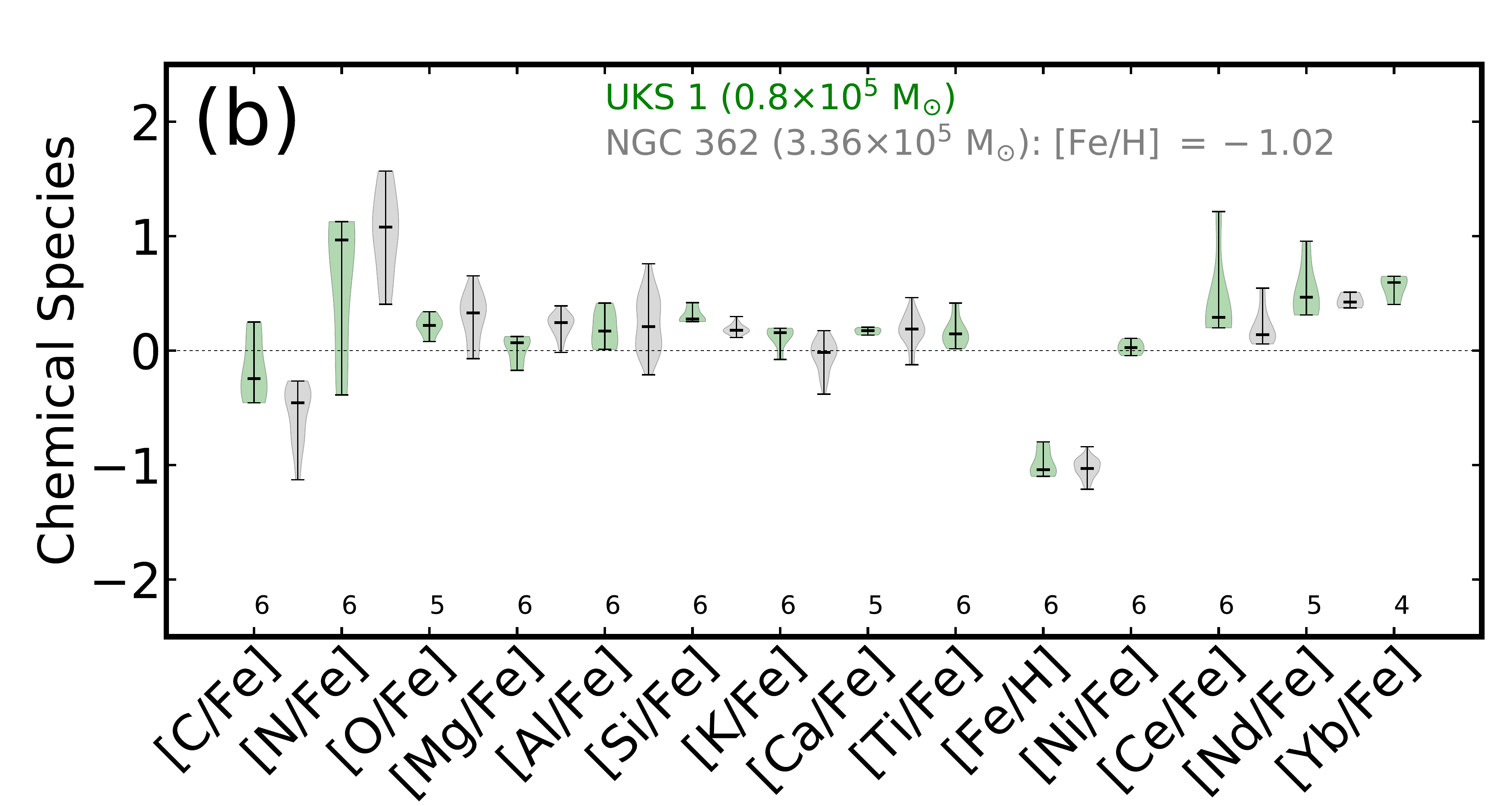}
     		\includegraphics[width=95mm]{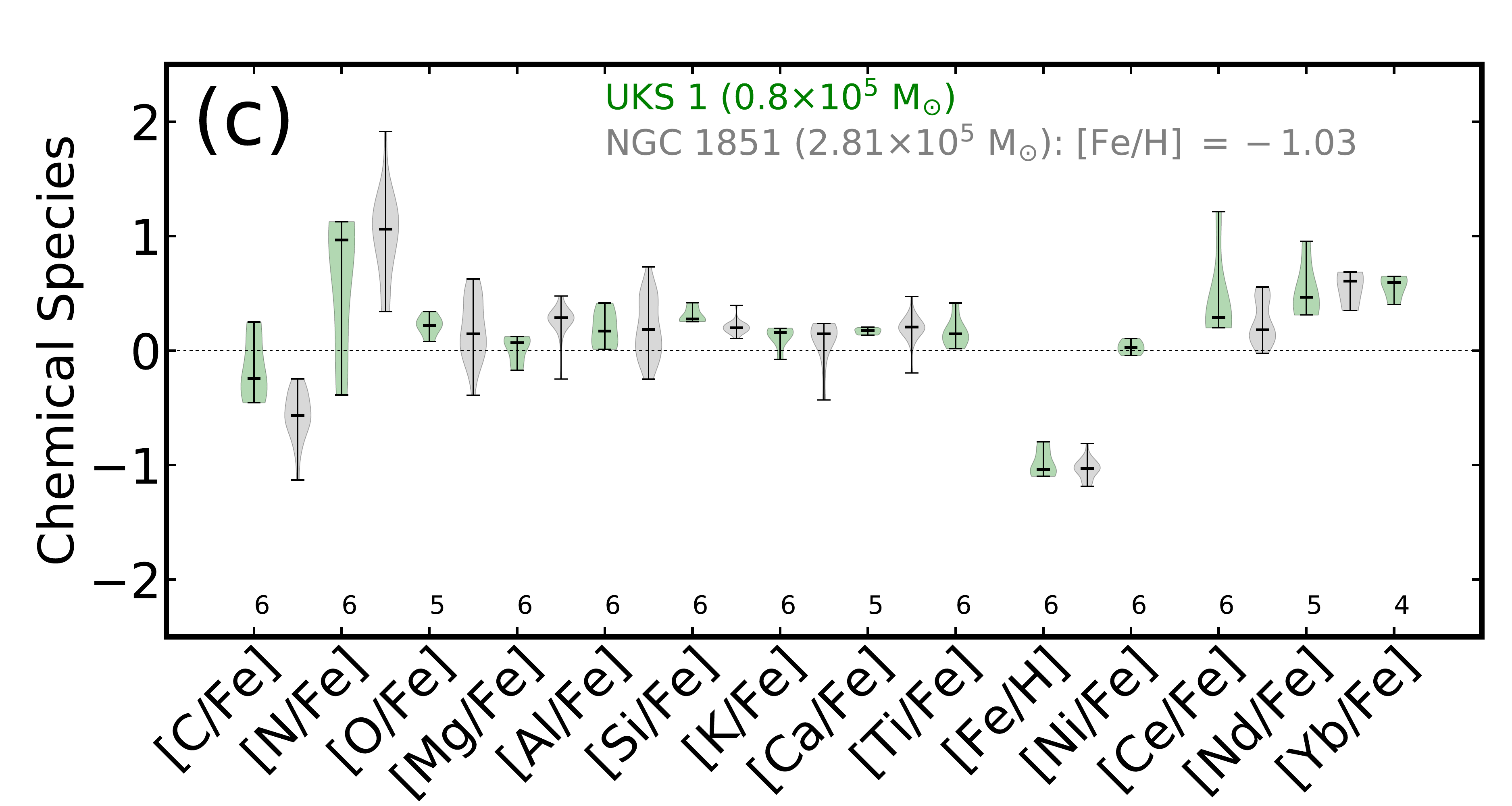}\includegraphics[width=95mm]{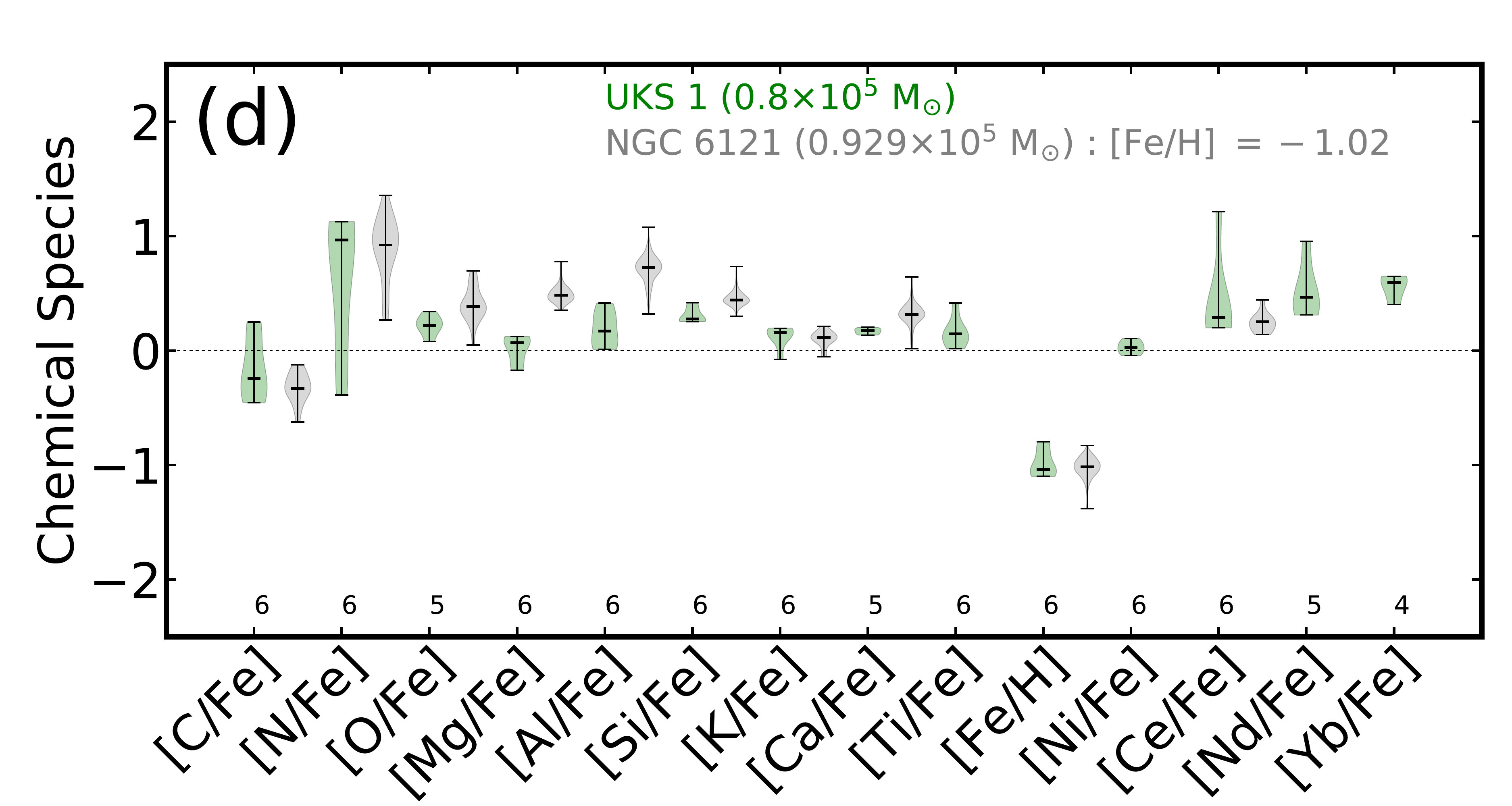}
     		\includegraphics[width=95mm]{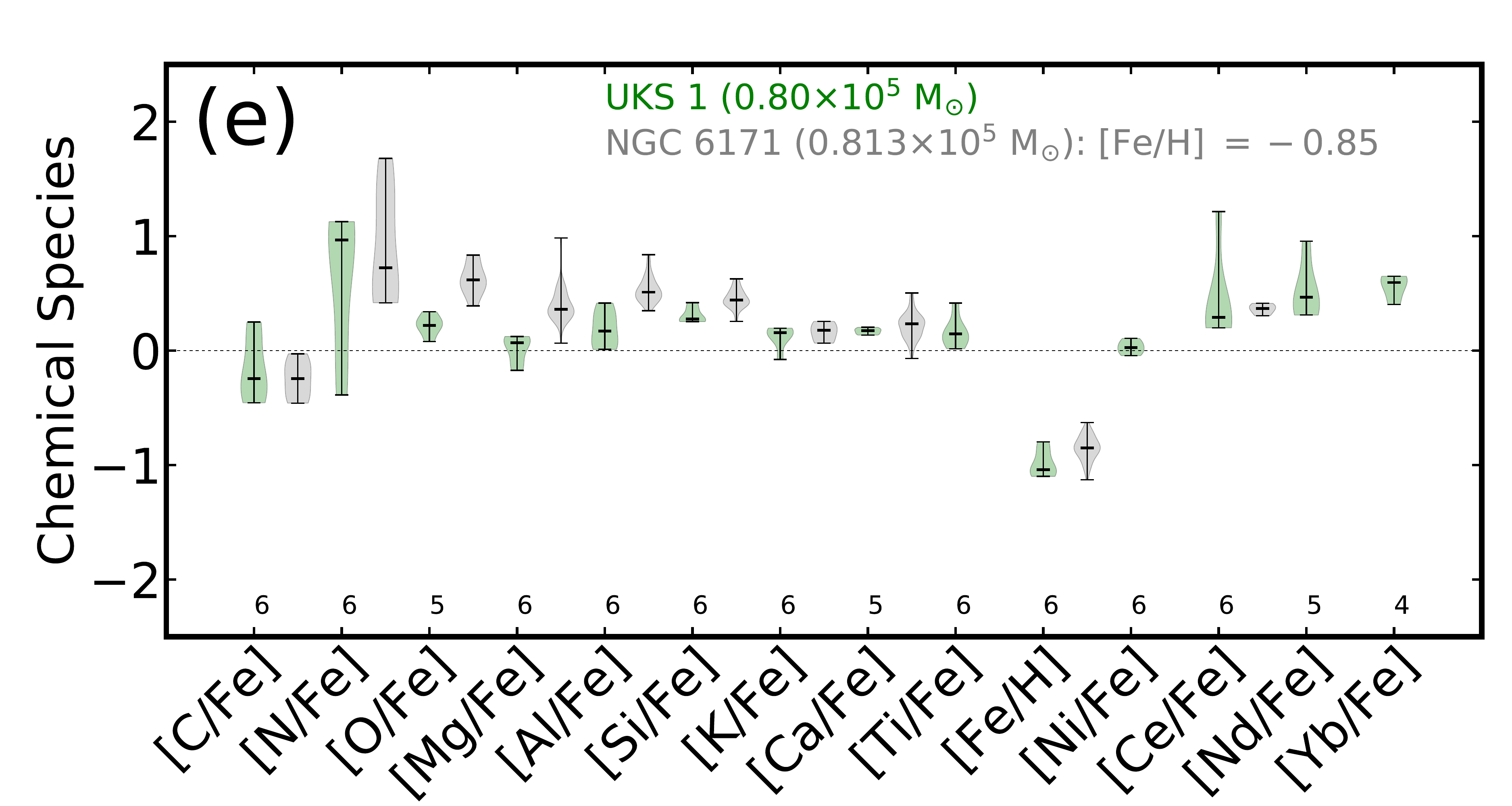}
     		\caption{[X/Fe] and [Fe/H] abundance density estimation comparison between UKS~1 (left green symbols) and GCs (right grey symbols), following the constraints outlined in Table 5 from \citet{Szabolcs2020}. Each violin representation indicates with horizontal lines the median and limits of the distribution. The corresponding number of stars with available abundances in our sample is marked in the bottom labels. The top label indicates the cluster mass from \citet{Baumgardt2019}.}
     		\label{Figure4}
     	\end{center}
     \end{figure*}	
     
      \subsection{The Fe-peak elements: Fe and Ni}
      
       From our overall spectral analysis we find an average [Fe/H]$=-0.98$, with scatter of $\sigma=$0.11 dex, and star-to-star [Fe/H] spread of 0.3 dex. This metallicity is on average $\approx 0.25$ dex lower than that reported in \citet{Origlia2005}, but with a difference not greater than our measured [Fe/H] spread. We also note that the large star-to-star [Fe/H] spread measured in this work is roughly comparable to the typical errors of [Fe/H] in some stars in our sample. Therefore, the large intrinsic [Fe/H] spread found in this work could be due to the large line-by-line (Fe I) variation in our spectra (see Table \ref{Table2}). 
      
        Some stars is our sample are as metal-rich ([Fe/H]$\sim$-0.79) as those found by \citet{Origlia2005}, but having found more metal-poor members of UKS~1, this naturally implies that our mean [Fe/H] for the cluster differs from theirs. Figure \ref{Figure4} shows that our [Fe/H] spread observed in UKS~1 is consistent with that seen in NGC 2808, NGC 1851, NGC 362, NGC 6171, and NGC 6121 \citep[see][]{Szabolcs2020}. We show that UKS~1 hosts stars as metal-poor as [Fe/H]$=-1.09$ and as metal rich as that of \citet{Origlia2005}. Based on our analysis, UKS~1 is likely a GC with an intermediate [Fe/H]$=-0.98$.
                  
        For nickel (Ni), UKS~1 has $\langle$[Ni/Fe]$\rangle =0.03\pm0.05$ (with a star-to-star spread $\sim$0.15 dex) similar to that observed in other GCs, for example NGC 6723 \citep{Crestani2019}, and is slightly enhanced, similar to bulge field stars at metallicities above [Fe/H]$\sim -$1 dex \citep[e.g.][]{Johnson2014, Bensby2017}.
         
      \subsection{The light-elements: C and N}
      
      We also measure an average carbon abundance in UKS~1 of $\langle$[C/Fe]$\rangle = -0.16\pm0.26$, which is slightly higher than that reported in \citet{Origlia2005}. Figure \ref{Figure4} shows that UKS~1 has a median distribution in agreement with the result by \citet{Szabolcs2020} for GCs, with a large star-to-star [C/Fe] spread ($\gtrsim$0.7 dex), higher than the scatter. We also find that the [C/Fe] spread seen in UKS~1 is comparable to that commonly seen in high-mass GCs at similar metallicity (see Figure \ref{Figure4}), but has a median distribution slightly higher than those commonly massive GCs. This [C/Fe]-mass trends could indicate that UKS~1 is as massive as NGC 6171 and NGC 6121, but with a significantly large [C/Fe] spread. 
      
      As can be seen in Table \ref{Table1}, there are two clear groups of stars in our sample. Two stars in UKS~1 exhibit low nitrogen abundances ([N/Fe]$\lesssim +0.03$)  with a slight enrichment in carbon ([C/Fe] ($\lesssim +0.24$), while the remaining four stars are strongly enriched in nitrogen ([N/Fe]$\gtrsim +0.9$) and accompanied by lower levels of carbon ([C/Fe]$<0$). A bimodality is clearly seen in [N/Fe] and [C/Fe], suggesting that UKS~1 exhibits a N-C anti-correlation (see Figure \ref{Figure8}(b)), which is clearly much larger than the typical errors in [C/Fe] and [N/Fe]. UKS~1 displays a significant [C/Fe] spread and a significant [N/Fe] spread, $>1$ dex and $\gtrsim0.7$dex, respectively, which is similar to that of other GCs, such as NGC 2808 \citep[see, e.g.][]{Szabolcs2020}. This is the first time that the presence of spread in [N/Fe] has been established for UKS~1.  We conclude that this high [N/Fe] abundance is indicative of the presence of multiple populations in UKS~1.
                    
      \subsection{The $\alpha$-elements: O, Mg, Si, Ca, and Ti}
      
      We have found mean values (star-to-star spread) for $\langle$[O/Fe]$\rangle = +0.22\pm0.08$ ($\sim0.26$ dex), $\langle$[Mg/Fe]$\rangle = +0.02\pm0.11$ ($\sim0.29$ dex), $\langle$[Si/Fe]$\rangle = +0.31\pm0.06$ ($\sim0.16$ dex), $\langle$[Ca/Fe]$\rangle = +0.17\pm0.03$ ($\sim0.06$ dex), and $\langle$[Ti/Fe]$\rangle = +0.16\pm0.12$ ($\sim0.39$ dex). Figure \ref{Figure4} shows that $\alpha$-element abundances are compatible with other Galactic GCs \citep{Szabolcs2020}, however, small differences of the median [O,Mg,Si,Ca,Ti/Fe] distributions can be drawn from Figure \ref{Figure4} as a function of the cluster mass. Our observed [O/Fe], [Mg/Fe], and [Si/Fe] abundances are slightly lower than that of low-mass GCs, but show comparable levels to high-mass GCs. On the other hand, [Ca/Fe] is in good agreement with GCs of similar metallicity (Figure \ref{Figure4}), with small star-to-star spread. Overall, UKS~1 displays slightly higher $\alpha-$element enhancement, a possible signature of the fast enrichment provided by supernovae (SNe) II events, as expected in old GCs \citep[see, e.g.][]{Crestani2019}.
    
     \citet{Origlia2005} successfully measured [O/Fe], [Mg/Fe], [Si/Fe], [Ca/Fe], and [Ti/Fe] abundances for UKS~1. Their measured abundances for these are consistent with the ones determined in this work. However, they find that [Mg/Fe] is enhanced ($\sim +0.3$) with small star-to-star spread ($<0.08$ dex), while we find low-Mg stars and slightly sub-solar [Mg/Fe] values with an average $\langle$ [Mg/Fe] $\rangle$ close to Solar, and a high star-to-star spread ($>0.29$ dex). We suspect that these two distinctive Mg populations in UKS~1 likely correspond to two differentiated stellar populations.
  
\begin{figure*}	
    	\begin{center}
    		\includegraphics[width=170mm]{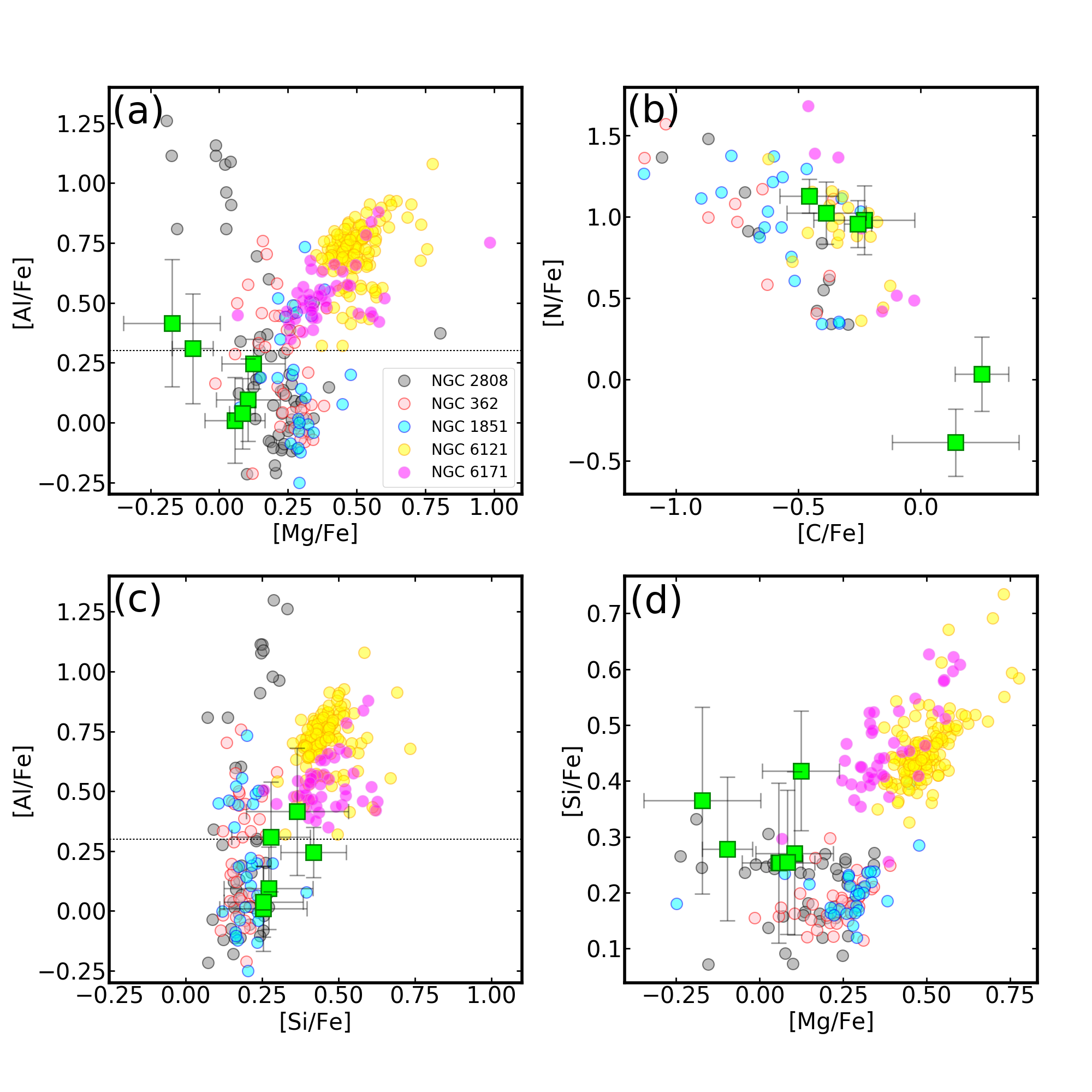}
    		\caption{Distributions of light- (C, N), $\alpha$- (Mg, Si) and odd-Z (Al) elements in different abundance planes: In panels (a), (b), (c), and (d) the planes [Al/Fe] -- [Mg/Fe], [N/Fe]--[C/Fe], [Al/Fe]--[Si/Fe], [Si/Fe]--[Mg/Fe] are respectively shown for GCs from \citet{Szabolcs2020}. The black dotted line at [A/Fe] $ = +0.3$ indicates the generalized separation of FG and SG stars as proposed in \citet{Szabolcs2020}. The distributions for UKS~1 star candidates (lime squares) are overlaid.}
    		\label{Figure8}
    	\end{center}
    \end{figure*}

     From Figure \ref{Figure8}(d)  we are able to confirm the presence of a weak Si-Mg anti-correlation in USK 1, similar to that seen in the massive cluster NGC 2808, indicating that leakage from the MgAl chain into Si production is also likely present in UKS~1. Figure \ref{Figure8}(c) also shows the presence of a weak Al-Si correlation, confirming the possible existence of $^{28}$Si leakage from the MgAl chain. Interestingly, most of the Si-enriched stars in UKS~1 also seem to correspond to the extreme Mg-depleted ([Mg/Fe]$<0$) and midly enhanced Al stars, suggesting that hot proton burning \citep[see, e.g.][]{Masseron2019, Szabolcs2020} in the early populations of UKS~1 may have taken place.
               
      It is also important to note that UKS~1 contains a fraction of stars well below [Mg/Fe]$\simeq-0.1$, therefore our finding of low-Mg stars is consistent with our lower metallicity for UKS~1. To our knowledge, sub-solar Mg abundances have not been found in GCs more metal rich than [Fe/H]$\sim-0.8$ \citep[see, e.g.][]{Meszaros2015, Pancino2017, Masseron2019, Szabolcs2020}.
   
       \subsection{The odd-Z elements: Al and K}
      
      Regarding Al and K, we found mean values for $\langle$[Al/Fe]$\rangle = +0.18\pm0.15$ and $\langle$[K/Fe]$\rangle = +0.12\pm0.09$, with a star-to-star spread of $\sim0.40$ dex and $\sim0.27$ dex, respectively. Figure \ref{Figure4} shows that the median distributions of these two chemical species are comparable to GCs at similar metallicity, except [Al/Fe], which exhibits a lower [Al/Fe] enrichment than that seen in low-mass GCs, but comparable to that of high-mass GCs.   
      
      Figure \ref{Figure8}(a) shows a clear anti-correlation between Al and Mg in UKS~1, which is larger than the typical errors of [Al/Fe] and [Mg/Fe] by a factor of $\sim$1.5, showing the signs of the Mg-Al cycle \citep[see, e.g.][]{Szabolcs2020}. By setting the [Al/Fe] limit at around 0.3 dex, as proposed by \citet{Szabolcs2020}, it is possible to roughly separate the so-called first (FG) and second (SG) generation stars. We identified 2 out of 6 stars with [Al/Fe]$\gtrsim +0.3$ (one of them strongly enriched in nitrogen), while the remaining stars in our sample exhibit a lower aluminium enrichment accompanied by a strong enrichment in nitrogen. This is similar to what is expected in intermediate-mass metal-rich asymptotic giant branch (AGB) stars, where N variations are expected without, or with very little, variation in Al \citep[see, e.g.][]{Ventura2013, Ventura2016}. The small Al production suggests that the Mg-Al cycle is modest in UKS~1, and seems compatible with that of the first stellar generation, similar to that observed in the massive accreted GC NGC 2808 \citep[see, e.g.][]{Szabolcs2020}. 
       
      Figure \ref{Figure4} suggests that UKS~1 displays a chemical enrichment similar to that of NGC 2808, which also shows a large spread in Al. Compared to other GCs, it is possible that UKS~1 has not been formed in the Milky Way, considering that Al was not abundant at the time of its formation and FG stars had lower Al than other clusters, as suggested by \citet[][]{Szabolcs2020} for the case of NGC 2808.   
     
      \subsection{The \textit{s}-process elements: Ce, Nd, and Yb}
       
      For the \textit{s}-process elements, our abundance analysis yields an average (star-to-star spread) $\langle$[Ce/Fe]$\rangle = 0.43\pm0.36$ ($\gtrsim$1 dex), $\langle$[Nd/Fe]$\rangle = 0.51\pm0.23$ ($\gtrsim$0.6 dex), and $\langle$[Yb/Fe]$\rangle = 0.56\pm0.10$ ($\gtrsim$0.25 dex). Figure \ref{Figure4} shows that UKS~1 displays an enrichment in \textit{s}-process elements comparable to other GCs at similar metallicity \citep{Szabolcs2020}, but with a large star-to-star scatter. It is possible that the modest enrichment in \textit{s}-process elements, accompanied by the lower levels in carbon and aluminium and the high enrichment in nitrogen, could tentatively be produced by intermediate-mass ($\sim$3 M$_{\odot}$) AGB stars \citep{Ventura2016}, supporting the pollution of this cluster by such stars. We can clearly see in Figure \ref{Figure2} that the majority of our stars lie in the very upper part of the RGB of UKS~1, which is in reasonable agreement with the expected behaviour for AGB stars. It is thus believed that a significant fraction of the stars in our sample are evolved AGB stars, possibly of intermediate masses.   
    
      \section{Orbit}
      \label{section6}
     
      We briefly investigate the implications of our PM measurements (see Appendix \ref{section8}) and the discrepancy in the heliocentric distance by calculating some possible orbits for UKS~1. The radial velocity adopted in this work is 66.13$\pm$12.89 km s$^{-1}$, which was obtained from the average of our stars from APOGEE with accurate radial velocity measurements.
      
      For the orbit computation, we adopt a sophisticated orbit modelling algorithm-- \texttt{GravPot16}\footnote{\url{https://gravpot.utinam.cnrs.fr}}, which considers the Galactic perturbations due to a physical `boxy/peanut' bar structure \citep[see, e.g.][]{Fernandez-Trincado2020}, and the superposition of ten disk components (including the ISM contribution) surrounded by an oblate Hernquist stellar and spherical dark matter halo, whose density profiles mimic that of the Besan\c{c}on Galaxy model \citep[][]{Robin2003, Robin2012, Robin2014}. For a more detailed description regarding the formalism of the \texttt{GravPot16} model, we refer the reader to a forthcoming paper (Fern\'andez-Trincado et al., in preparation). 
      
\begin{figure*}	
 	\begin{center}
 		\includegraphics[width=90mm]{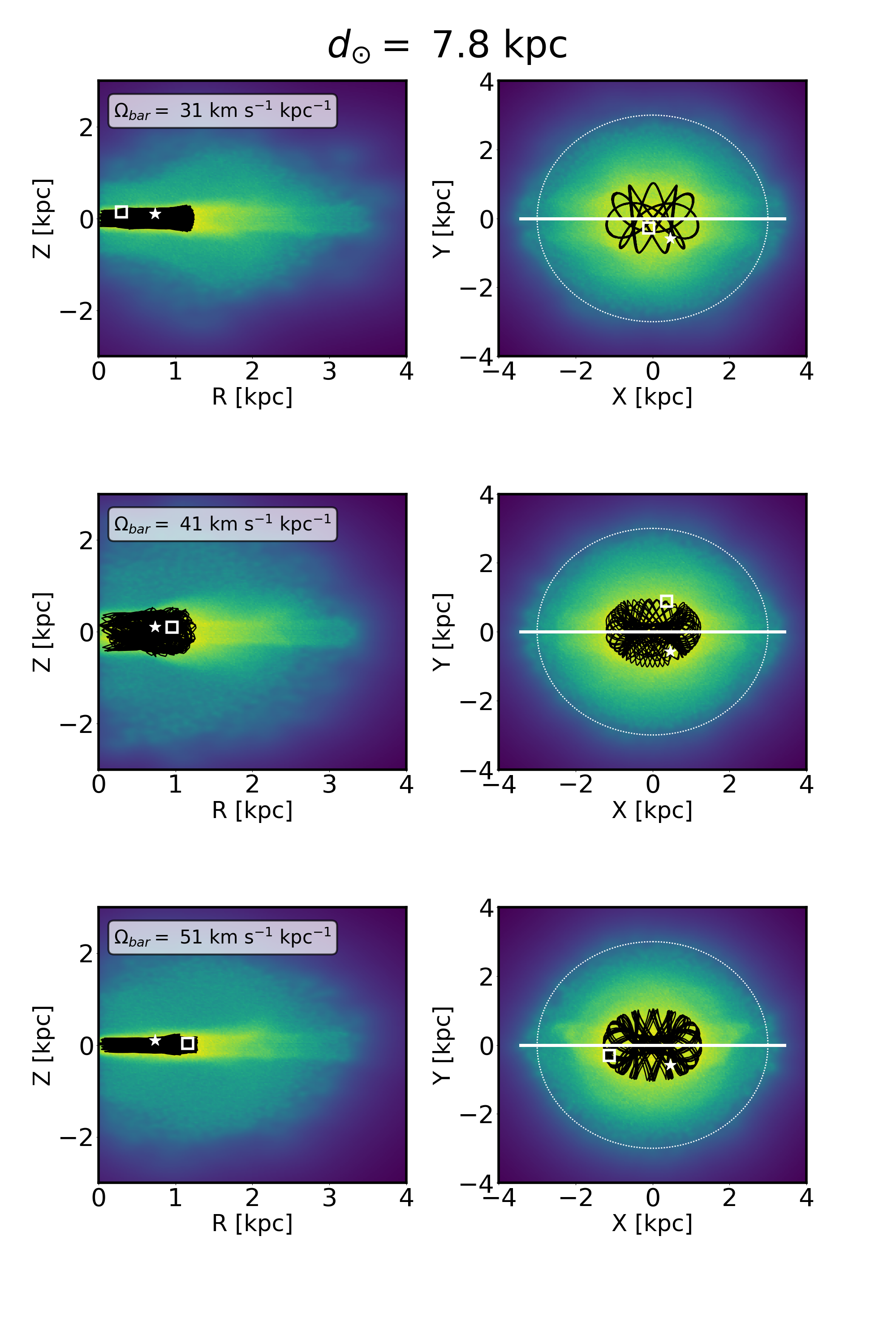}\includegraphics[width=90mm]{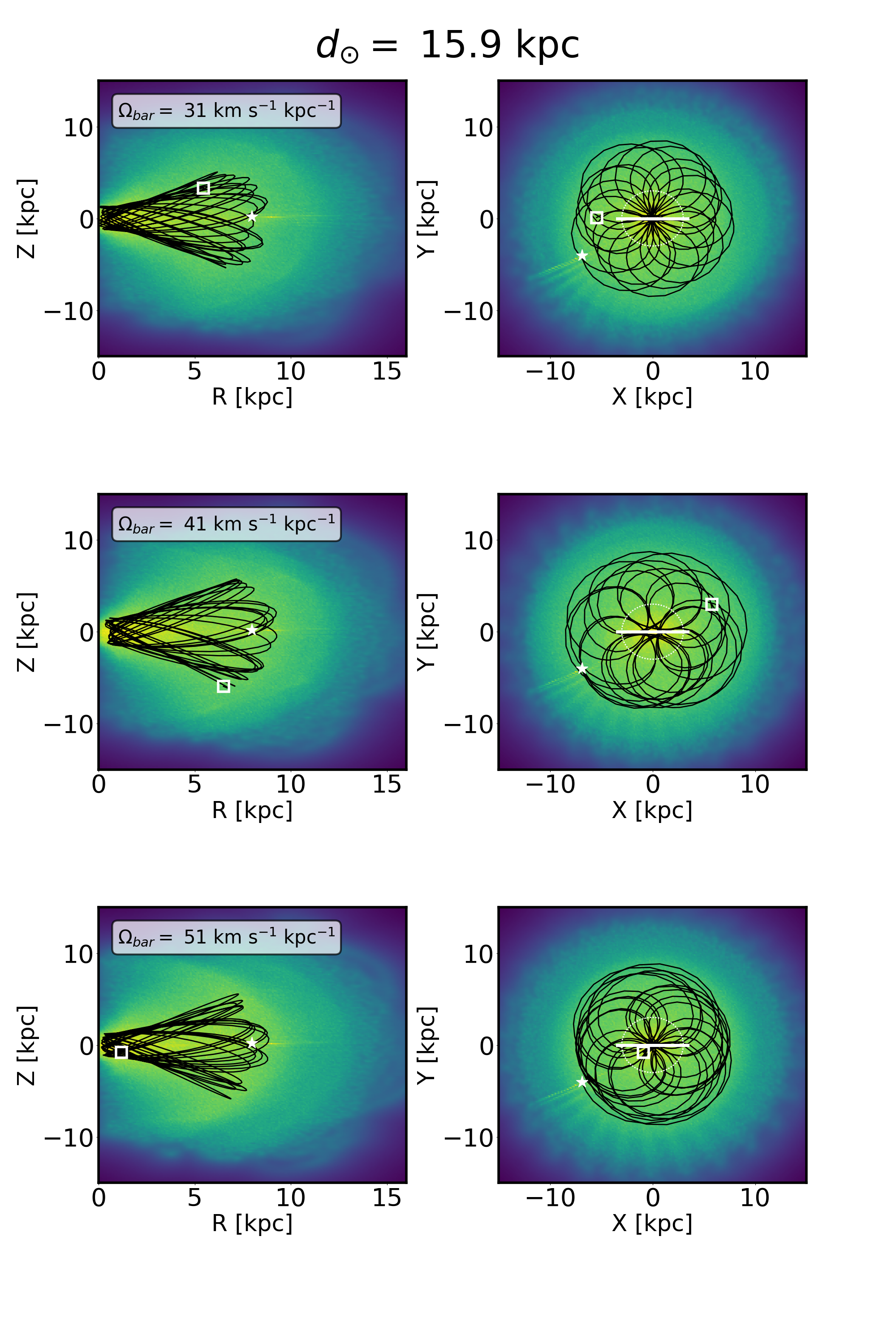}
 		\caption{Probability density map for the \textit{R}-\textit{z} and \textit{x}-\textit{y} projections of a an ensemble of 100,000 orbits for UKS~1 in the reference frame where the bar is at rest. Simulations for three different value of the bar patterns speed ($\Omega_{\rm bar}$) are shown. The white dotted circle indicates the assumed bulge radius of $\sim3$ kpc according to \citet{Barbuy2018}; the thick white line shows the length ($\sim3.4$ kpc) of the physical boxy/peanut bar structure employed in the \texttt{GravPot16} model. The white star and square symbols indicate the initial and end position of the cluster after a 2 Gyr backwards integration time. The black lines show the orbit of UKS~1 by assuming the central values of the observables, while the yellow colour map show the more probable regions of the space crossed by the cluster (considering the uncertainty in the observables). The orbit results are presented by adopting two different heliocentric estimates, i.e. 7.8 kpc (\textit{left panels}) and 15.9 kpc (\textit{right panels}). The white `star' and square symbols indicate the present and final position of the orbit of the cluster, respectively.}
 		\label{Figure5}
 	\end{center}
 \end{figure*}

      In this study, we employed the same Galactic configuration and Solar motion defined in \citet[][]{Fernandez-Trincado2020}, except for the bar pattern speeds, for which we have assumed an angular velocity of the bar of $\Omega_{\rm bar}=41$ km s$^{-1}$ kpc$^{-1}$ \citep[see][]{Bovy2019}. Simulations were also obtained by adopting an uncertainty in $\Omega_{\rm bar}$ of $\pm10$ km s$^{-1}$ kpc$^{-1}$. The Galactic potential has been rescaled to the Sun's galactocentric distance,  $R_{\odot}=8$ kpc, and the local rotation velocity, $\Theta_{\rm 0} = 244.5$ km s$^{-1}$, given by \citet{Robin2017}. To place error bars on the orbital elements, we integrate an ensemble of 100,000 orbits, randomly selecting values from Gaussian distributions centred on the mean values of distance, proper motions, radial velocity, and Galactic parameter variations. Figure \ref{Figure5} shows this ensemble of orbits as yellow/green-coloured paths, for two different heliocentric distances from \citet{Minniti2011} and \citet{Baumgardt2019}. Table \ref{Table3} lists the orbital parameters and their uncertainties, which have been estimated as the 50$^{\rm th}$ (median), 16$^{\rm th}$, and 84$^{\rm th}$ percentiles of the distributions resulting from integrations over a 2 Gyr timespan.
           
     Figure \ref{Figure5} shows that UKS~1 is bound to the `bar/bulge' region ($r_{apo}\lesssim1.4$ kpc and $|Z_{max}|\lesssim0.5$ kpc) in a radial and rather eccentric ($\gtrsim0.97$) P-R\footnote{We call prograde-retrograde (P-R) orbits the ones that flip their sense from prograde to retrograde, or vice versa, along their orbits.} orbit, assuming a heliocentric distance of $\sim$7.8 kpc \citep{Baumgardt2019} with 10\% of uncertainty. It is important to note that our measured orbital properties for UKS~1 are similar to those reported by \citet{Baumgardt2019}, despite the fact that there is a small difference in proper motions. Therefore, our measured proper motions with VIRAC have no dramatic effect on the orbital configuration. 
     
     A different picture is obtained if UKS~1 is initially positioned behind the Galactic bulge at $\sim$15.9 kpc \citep{Minniti2011}, displaying in this case radial and high eccentric ($e \gtrsim0.94$) prograde or P-R orbits (depending on the bar's angular velocity) confined to the inner halo ($|Z_{max}|\lesssim6$ kpc), within $\lesssim$9 kpc of the Galactic centre. This is contrary to the findings of \citet{Baumgardt2019}. If UKS~1 is confimed to be a distant GC, it would be a possible candidate to be associated with the \textit{Gaia}-Enceladus-Sausage early merger event \citep[][]{Belokurov2018}, with similar chemistry and orbital properties as NGC 2808 \citep{Baumgardt2019, Massari2019}. 
     
\begin{figure*}	
 	\begin{center}
 		\includegraphics[width=190mm]{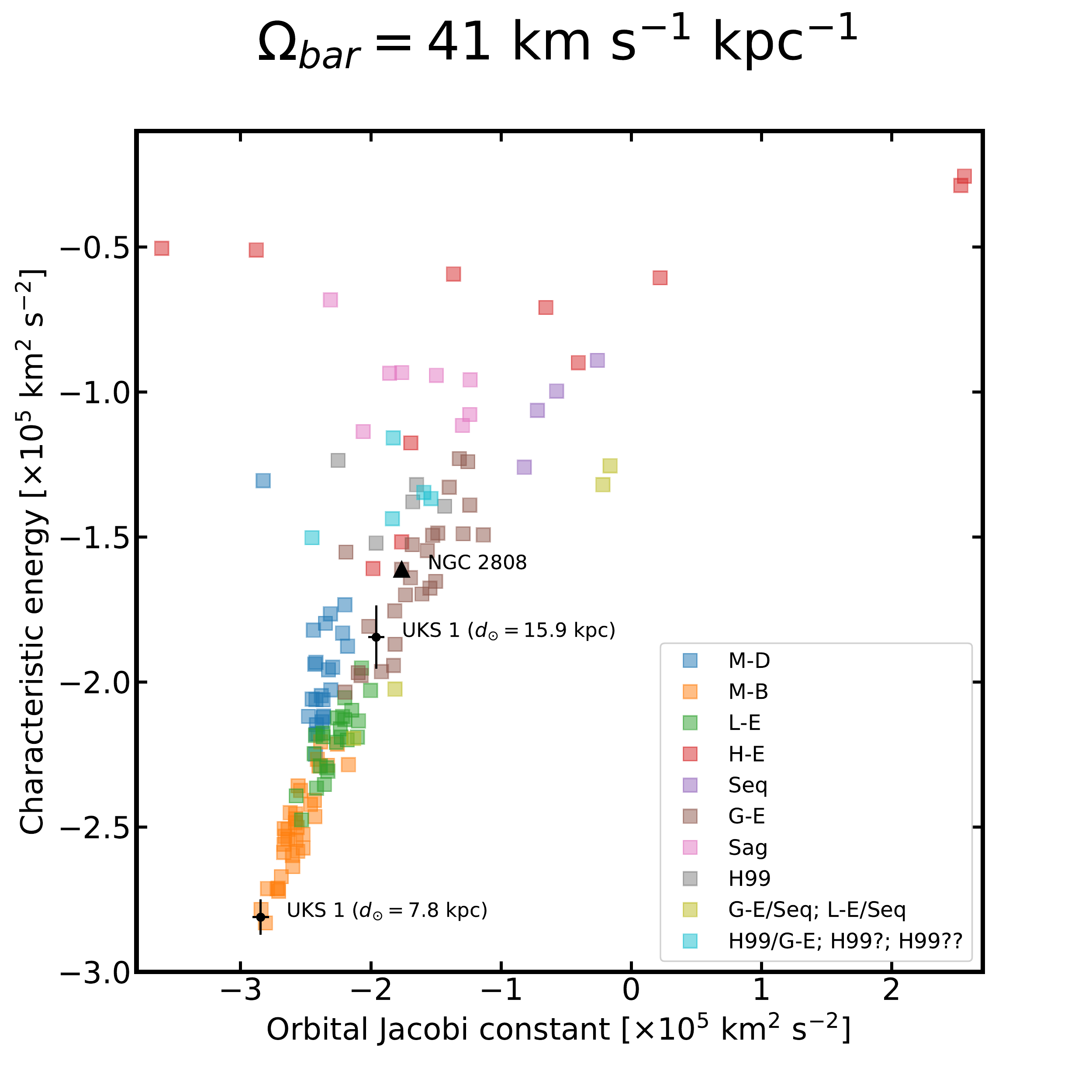}
 		\caption{Characteristic orbital energy ($(E_{max}+E_{min})/2$) versus the orbital Jacobi constant ($E_{J}$) in the non-inertial reference frame where the bar is at rest. Square symbols refer to Galactic GCs, colour-coded according to their associations with different progenitors from \citet{Massari2019}. The revisited dynamics of UKS~1, adopting the input parameters presented in this work, is shown with the black crosses. NGC 2808 (black triangle) is highlighted for reference.}
 		\label{Figure6}
 	\end{center}
 \end{figure*}

     Figure \ref{Figure6} follows these two possible scenarios. This figure shows the characteristic orbital energy versus the orbital Jacobi constant as envisioned by \citet{Moreno2015} and \citet{Fernandez-Trincado2020}, which is conserved in the reference frame where the bar is at rest. We can clearly see the two possible scenarios for UKS~1: (i) assuming a distance of 7.8 kpc leads to the conclusion that the cluster belongs to the GC family trapped into the bulge and possibly in the bulge/bar region; (ii) assuming a distance of 15.9 kpc positions UKS~1 in the region dominated by the group of GCs thought to belong to the \textit{Gaia}-Enceladus-Sausage galaxy merger debris, such as NGC 2808.
           
     We notice that the properties of the Galatic potential affect the orbits only to a lesser degree. Variations in the heliocentric distance completely dominates the uncertainty in our understanding of the global dynamical picture of UKS~1.

\section{Age}
\label{section10}

Deriving the age for the UKS~1 GC is not an easy task because, as mentioned before, it is in a region with very high extinction \citep[see, e.g.][]{Minniti2011}.  Through isochrone fitting, we try to provide a rough estimation of the age. To accomplish that, we employ the SIRIUS code \citep{Souza2020}, which applies a statistical Bayesian Markov Chain Monte Carlo method. 

For the isochrone fitting, we adopt the Darthmouth Stellar Evolutionary Database \citep[DSED;][]{Dotter2008} with an $\alpha$-enhancement of $+0.4$ and canonical helium (Y$\sim0.25$) models. The DSED isochrones are available in the 2MASS photometry system, and they were converted to the VVV photometry system.

Since we do not have the entire CMD available, in particular, the turn-off region, we imposed Gaussian distribution priors for the metallicity of [Fe/H]$=-0.98$, based on the mean determination from this work, with a standard deviation of 0.11 dex, and for the distance of $7.8$ kpc \citep{Baumgardt2019} with a standard deviation of $0.78$ kpc. In contrast, we also employed a heliocentric distance of 15.9 kpc \citep{Minniti2011}. However, no age distribution converged.

\begin{figure*}	
	\begin{center}
		\includegraphics[width=69mm]{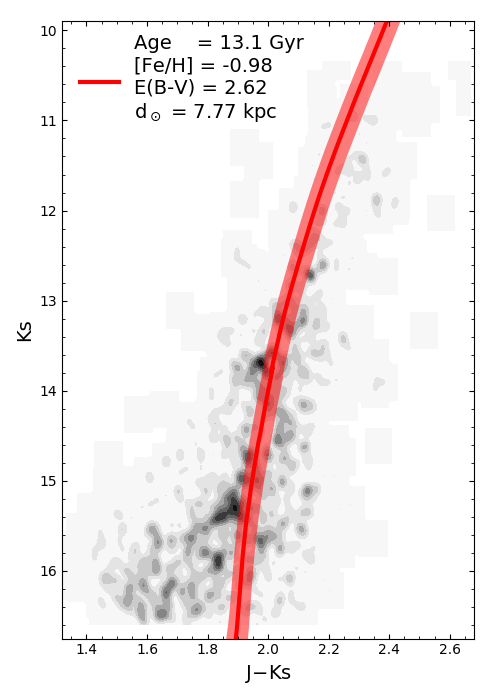}\includegraphics[width=95mm]{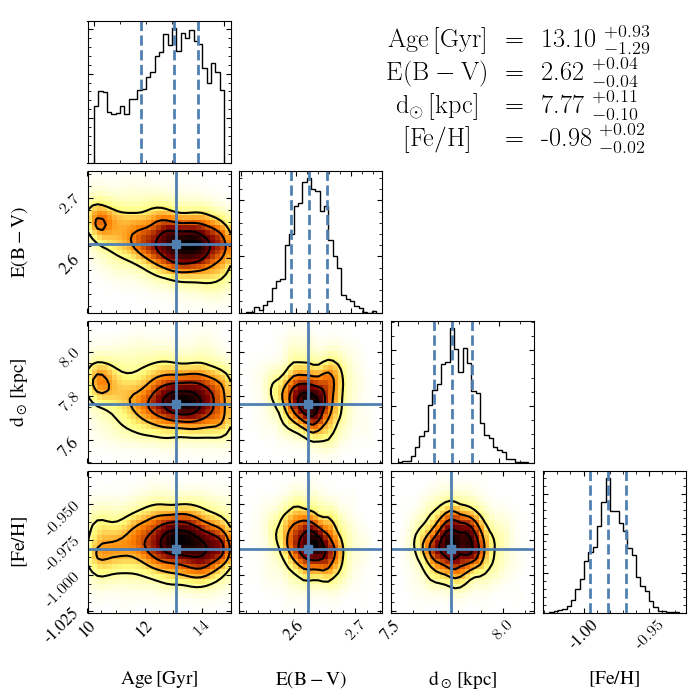}
		\caption{Best isochrone fit in the Ks versus $(J-K_{s})$ CMD using DSED models. \textit{Left panel:} CMD with the results from the fitting. The red line is the most probable solution and the red stripe are the solutions within $1-\sigma$. \textit{Right panel:} Posterior distributions.}
		\label{Figure7}
	\end{center}
\end{figure*}

Figure \ref{Figure7} presents the best isochrone fitting in the Ks versus (J-Ks) CMD. Our fit provides a reasonable solution both in the overplotted isochrone (left panel) and the posterior distributions of the corner plot (right panel).  As the best determination to represent the distributions, we adopted the median as the most probable value and the uncertainties calculated from 16th and 84th percentiles. Based on the DSED isochrones, we found an age of $13.10^{+0.90}_{-1.30}$ Gyr. 

The  $1-\sigma$ region (the red stripe in Figure \ref{Figure7}) is mostly affected by age uncertainties. It is relevant to mention that in the RGB region of the CMD, an age variation could be seen as a colour displacement \citep[see Figure 2 of ][]{Souza2020}. Also, we want to highlight that our probable solutions within $1-\sigma$ fit well the central part of the CMD, reinforcing that the age estimation is a reasonable determination for UKS~1. 

      \section{Concluding remarks} 
      \label{section7}
     	
     	We have employed the public APOGEE DR16 catalogue in combination with the internal release of the VIRAC catalogue to extensively characterize the chemical composition of UKS~1. We have performed a high-resolution spectral analysis of six stars in the innermost region of the cluster. Our main conclusions are summarized as follows:
     	
     	\begin{itemize}
     		
     		\item[$\bullet$]  We find an intermediate metallicity, [Fe/H]$=-0.98$, with a star-to-star spread of 0.11 dex. The reported metallicity is $\sim$0.25 dex more metal-poor that previously reported in \citet{Origlia2005} with a smaller sample size (3 stars).  
     		
     		\item[$\bullet$] We find a radial velocity of 66.13 km s$^{-1}$, which is in good agreement with \citet{Baumgardt2019}. However, the radial velocity dispersion is found to be slightly higher, $\sim$12.89 km s$^{-1}$, but still typical of some GCs \citep[see, e.g.][]{Baumgardt2018}. Due to the observed nitrogen over-abundance (see below) in our sample, the possibility of these stars being field stars is very low. 
     		
     		\item[$\bullet$] We identified four stars in the innermost region of UKS~1 with stellar atmospheres strongly enriched in nitrogen ([N/Fe]$>+0.95$) accompanied by low carbon abundance ratios ([C/Fe]$\lesssim-0.23$). There is clear evidence for a bimodal distribution in the abundances of C and N present in our data, exhibiting a clear C-N anti-correlation. This result indicates the prevalence of the multiple-population phenomenon in UKS~1.
     		
     		\item[$\bullet$] Even though our analysis strongly supports the hypothesis that UKS~1 is a GC native to the bulge region, better constraints in the distance will help confirm or refute if it is a permanent resident of the Galactic bulge or the inner stellar halo. 
     		
     	    \item[$\bullet$] We provide, for the first time, a rough estimate of the age for UKS~1 by adopting a statistical isochrone fitting technique in the VVV CMDs following a Bayesian approach, which determine the ages, distances, metallicity, and reddening in a self-consistent way \citep[see, e.g.][]{Souza2020}. We assumed that UKS~1 currently lies in the bulge, which returns an age estimated of 13.10$^{+0.93}_{-1.29}$ Gyr (see Section \ref{section10}), suggesting that UKS~1 is an old cluster in the Galactic bulge. We attempted to obtain consistent results by assuming a large distance ($\sim$15.9 kpc); this proved unsuccessful. With deeper photometry, we could confirm the age estimation and distance that would be in favour of the scenario that UKS~1 is located in the bulge, as suggested by the distance reported in the Baumgardt's catalogue \citep{Baumgardt2018, Baumgardt2019, Hilker2020}.
     	    
     	\end{itemize}

	\begin{acknowledgements}  
	The authors thank the expert anonymous referee, who provided generous detailed feedback that substantially improved the paper. We gratefully acknowledge data from the ESO Public Survey program ID 179.B-2002 taken with the VISTA telescope, and products from the Cambridge Astronomical Survey Unit (CASU) and from the VISTA Science Archive (VSA). This publication makes use of data products from the WISE satellite, which is a joint project of the University of California, Los Angeles, and the Jet Propulsion Laboratory/California Institute of Technology, funded by the National Aeronautics and Space Administration. This research has made use of NASAs Astrophysics Data System Bibliographic Services and the SIMBAD database operated at CDS, Strasbourg, France.\\
     
	J.G.F-T is supported by FONDECYT No. 3180210. D.M. is supported by the BASAL Center for Astrophysics and Associated Technologies (CATA) through grant AFB 170002, and by project FONDECYT Regular No. 1170121. D.G. gratefully acknowledges support from the Chilean Centro de Excelencia en Astrof\'isica y Tecnolog\'ias Afines (CATA) BASAL grant AFB-170002. D.G. also acknowledges financial support from the Direcci\'on de Investigaci\'on y Desarrollo de la Universidad de La Serena through the Programa de Incentivo a la Investigaci\'on de Acad\'emicos (PIA-DIDULS). T.C.B. acknowledges partial support for this work from grant PHY 14-30152: Physics Frontier Center / JINA Center for the Evolution of the Elements (JINA-CEE), awarded by the US National Science Foundation. S.V. gratefully acknowledges the support provided by Fondecyt reg. n. 1170518. J.A.-G. acknowledges support from Fondecyt Regular 1201490 and from ANID, Millennium Science Initiative ICN12\_009, awarded to the Millennium Institute of Astrophysics (MAS). S.O.S acknowledges the FAPESP PhD fellowship 2018/22044-3. A.P-V acknowledges the FAPESP postdoctoral fellowship no. 2017/15893-1 and DGAPA-PAPIIT grant IG100319. B.B acknowledge partial financial support from FAPESP, CNPq, and CAPES - Finance Code 001. B.T. gratefully acknowledges support from National Natural Science Foundation of China under grant No. U1931102 and support from the hundred-talent project of Sun Yat-sen University.\\
	
	\texttt{BACCHUS} have been executed on the Supercomputer TITAN from the Departamento de Astronom\'ia de la Universidad de Concepci\'on.\\
	
	Funding for the Sloan Digital Sky Survey IV has been provided by the Alfred P. Sloan Foundation, the U.S. Department of Energy Office of Science, and the Participating Institutions. SDSS- IV acknowledges support and resources from the Center for High-Performance Computing at the University of Utah. The SDSS web site is www.sdss.org. SDSS-IV is managed by the Astrophysical Research Consortium for the Participating Institutions of the SDSS Collaboration including the Brazilian Participation Group, the Carnegie Institution for Science, Carnegie Mellon University, the Chilean Participation Group, the French Participation Group, Harvard-Smithsonian Center for Astrophysics, Instituto de Astrof\`{i}sica de Canarias, The Johns Hopkins University, Kavli Institute for the Physics and Mathematics of the Universe (IPMU) / University of Tokyo, Lawrence Berkeley National Laboratory, Leibniz Institut f\"{u}r Astrophysik Potsdam (AIP), Max-Planck-Institut f\"{u}r Astronomie (MPIA Heidelberg), Max-Planck-Institut f\"{u}r Astrophysik (MPA Garching), Max-Planck-Institut f\"{u}r Extraterrestrische Physik (MPE), National Astronomical Observatory of China, New Mexico State University, New York University, University of Notre Dame, Observat\'{o}rio Nacional / MCTI, The Ohio State University, Pennsylvania State University, Shanghai Astronomical Observatory, United Kingdom Participation Group, Universidad Nacional Aut\'{o}noma de M\'{e}xico, University of Arizona, University of Colorado Boulder, University of Oxford, University of Portsmouth, University of Utah, University of Virginia, University of Washington, University of Wisconsin, Vanderbilt University, and Yale University.\\
	
	\end{acknowledgements}
	

		\begin{sidewaystable*}
		\begin{small}
			\setlength{\tabcolsep}{0.50mm}  
			\caption{Basic parameters of Stars in UKS~1.}
			\centering
			\begin{tabular}{|c|cccccccccccccccc|}
				\hline
				Star Ids&  APOGEE\_ID   &  [C/Fe]  &   [N/Fe]     &  [O/Fe]    &  [Mg/Fe]      &  [Al/Fe]    & [Si/Fe]   &  [K/Fe]      &  [Ca/Fe]  & [Ti/Fe]     &    [Fe/H]   &  [Ni/Fe]    &  [Ce/Fe]  &  [Nd/Fe] &  [Yb/Fe] &  \\
				&     & &        &        &    &        &      & &      &    &     &     &     &  &   &    \\
				\hline
				\hline
1  &  2M17542167$-$2409501 &  $-$0.45  &  $+$1.13  &  $+$0.24 &  $+$0.10 & $+$0.09 & $+$0.27 & $-$0.07 & $+$0.13 & $+$0.139  & $-$1.06 & $-$0.02 & $+$0.22 & $+$0.51     &  $+$0.40 &  \\
2  &  2M17542506$-$2406036 &  $-$0.38  &  $+$1.02  &  $+$0.21 &  $+$0.05 & $+$0.01 & $+$0.25 & $+$0.18 & $+$0.17 & $+$0.158  & $-$1.02 & $-$0.04 & $+$0.26 & $+$0.32     &  $+$0.63 &  \\
3  &  2M17542652$-$2408478 &  $-$0.23  &  $+$0.99  &  $+$0.07 &  $-$0.09 & $+$0.31 & $+$0.27 & $+$0.15 & $+$0.14 & $+$0.016  & $-$1.09 & $+$0.07 & $+$0.31 & $+$0.95     &  ...     &  \\
4  &  2M17544067$-$2410464 &  $-$0.25  &  $+$0.95  &  $+$0.21 &  $+$0.08 & $+$0.04 & $+$0.25 & $+$0.11 & $+$0.20 & $+$0.151  & $-$1.05 & $+$0.02 & $+$0.33 & $+$0.31     &  $+$0.55 &  \\
5  &  2M17543085$-$2407438 &  $+$0.24  &  $+$0.03  &  $+$0.33 &  $+$0.12 & $+$0.25 & $+$0.41 & $+$0.19 & $+$0.18 & $+$0.076  & $-$0.79 & $+$0.10 & $+$0.19 & $+$0.46     &  $+$0.65 &  \\
6  &  2M17542434$-$2407356 &  $+$0.14  &  $-$0.39  &  ...     &  $-$0.17 & $+$0.42 & $+$0.36 & $+$0.15 & ...     & $+$0.415  & $-$0.87 & $+$0.03 & $+$1.21 & ...         &  ...     &  \\
				\hline
				\hline
				Star Ids&  APOGEE\_ID  &  $\alpha$&  $\delta$ &  T$_{\rm eff}$  & $\log$ \textit{g} & [M/H] & $\xi$ &  S/N & J$_{\rm 2MASS}$ & K$_{\rm s,2MASS}$ & ${\rm J-K}_{\rm s}^{*}$ & K$_{\rm s}^{*}$ & RV & $\sigma$RV  & $\mu_{\alpha}\cos(\delta)$ & $\mu_{\delta}$  \\ 
				&     & (hh:mm:ss) & (dd:mm:ss) &  (K)  &  & & (km s$^{-1}$)  & (pixel$^{-1}$) & (mag) & (mag) & (mag) & (mag) & (km s$^{-1}$) & (km s$^{-1}$)  & (mas yr$^{-1}$) & (mas yr$^{-1}$)  \\ 
				\hline
1  &  2M17542167$-$2409501  & 17:54:21.67 & $-$24:09:50.1 & 3756       & 0.23      &  $-$1.18 & 2.61 &  192 & 12.92 & 10.31   & 2.64$\pm$0.17 & 10.35$\pm$0.13  &   89.4 & 0.65     &   $-$4.01$\pm$1.04 &  $-$0.03$\pm$1.10  \\
2  &  2M17542506$-$2406036  & 17:54:25.06 & $-$24:06:03.6 & 3912       & 0.82      &  $-$1.04 & 2.86 &  148 & 13.28 & 10.81   & 2.44$\pm$0.06 & 10.79$\pm$0.04  &   70.7 & 0.36     &   $-$2.87$\pm$0.33 &  $-$2.63$\pm$0.37  \\
3  &  2M17542652$-$2408478  & 17:54:26.52 & $-$24:08:47.8 & 4001       & 1.12      &  $-$1.19 & 2.97 &  96  & 12.15 &  9.70   & 2.08$\pm$0.04 & 12.84$\pm$0.02  &   63.2 & ...      &   $-$0.68$\pm$0.84 &  $-$3.91$\pm$0.89  \\
4  &  2M17544067$-$2410464  & 17:54:40.67 & $-$24:10:46.4 & 3872       & 0.60      &  $-$1.04 & 2.64 &  134 & 13.16 & 10.64   & 2.67$\pm$0.15 & 10.75$\pm$0.11  &   64.2 & 0.16     &   $-$2.48$\pm$0.52 &  $-$2.05$\pm$0.57  \\
5  &  2M17543085$-$2407438  & 17:54:30.85 & $-$24:07:43.8 & 4146       & 1.70      &  $-$0.70 & 2.06 &  144 & 13.44 & 11.12   & 2.20$\pm$0.08 & 11.04$\pm$0.06  &   63.0 & 0.03     &   $-$5.07$\pm$0.24 &  $-$2.37$\pm$0.27  \\
6  &  2M17542434$-$2407356  & 17:54:24.34 & $-$24:07:35.6 & 4129       & 2.11      &  $-$0.88 & 2.94 &  182 & 13.29 & 10.89   & 2.49$\pm$0.09 & 10.97$\pm$0.07  &   45.9 & 1.92     &   $-$2.22$\pm$0.58 &  $-$3.00$\pm$0.61  \\
				\hline			
			\end{tabular}  \label{Table1}
				\tablefoot{($^{*}$) Denotes the differential reddening corrected K$_{\rm s}$ and ${\rm J-K}_{\rm s}$.}
		\end{small}
	\end{sidewaystable*}

     	\begin{table*}
	\begin{small}
		\begin{center}
			\setlength{\tabcolsep}{1.5mm}  
			\caption{Sensitivity to typical uncertainties in atmospheric parameters and standard deviation between lines of the same species. Iron lines represent the [Fe I/H] values, while all other species are the [X/Fe] ratios, for elements X $=$ C, N, O, Mg, Al, Si, K, Ca, Ti, Ni, Ce, Nd, and Yb.}
			\centering
			\begin{tabular}{|ccccccccccccccc|}
				\hline
				APOGEE\_ID       &    C   &    N   &    O   &    Mg   &    Al  &    Si  &    K   &    Ca  &    Ti  &    Fe  &    Ni  &    Ce  &    Nd  &    Yb  \\
				       &    (dex)   &    (dex)   &    (dex)      &    (dex)      &    (dex)     &   (dex)     &    (dex)      &    (dex)     &    (dex)     &    (dex)   &    (dex)     &    (dex)    &    (dex)     &  (dex)   \\
				\hline
				\hline
				2M17542167$-$2409501 & & & & & &  & &  & & &  & & & \\
			$\sigma_{\rm T_{\rm eff}}$	&  0.052 &  0.063 &  0.132 &  0.047  &  0.059 &  0.078 &  0.106 &  0.078 &  0.078 &  0.073 &  0.087 &  0.048 &  0.156 &  0.147 \\
			$\sigma_{\rm [X/H], \log \textit{g}}$	&  0.035 &  0.007 &  0.019 &  0.015  &  0.035 &  0.022 &  0.023 &  0.031 &  0.024 &  0.046 &  0.019 &  0.045 &  0.202 &  0.144 \\
			$\sigma_{\rm \xi_t}$	&  0.014 &  0.014 &  0.002 &  0.032  &  0.008 &  0.022 &  0.093 &  0.023 &  0.009 &  0.046 &  0.032 &  0.016 &  0.144 &  0.006 \\
			$\sigma_{\rm mean}$		&  0.101 &  0.080 &  0.154 &  0.100  &  0.157 &  0.119 &  0.199 &  0.070 &  0.088 &  0.088 &  0.193 &  0.059 &  0.080 &  ...   \\
			$\sigma_{\rm total}$		&  0.119 &  0.103 &  0.204 &  0.116  &  0.172 &  0.146 &  0.245 &  0.112 &  0.120 &  0.132 &  0.215 &  0.089 &  0.304 &  0.206 \\
				\hline
				\hline
				2M17542506$-$2406036 & & & & & &  & &  & & &  & & & \\
				$\sigma_{\rm T_{\rm eff}}$&  0.070 &  0.136 &  0.173 &  0.043  &  0.065 &  0.024 &  0.041 &  0.062 &  0.065 &  0.016 &  0.053 &  0.051 &  0.061 &  0.049 \\
				$\sigma_{\rm \log \textit{g}}$ &  0.117 &  0.082 &  0.037 &  0.077  &  0.070 &  0.006 &  0.008 &  0.014 &  0.026 &  0.009 &  0.024 &  0.048 &  0.080 &  0.125 \\
				$\sigma_{\rm \xi_t}$	&  0.070 &  0.039 &  0.014 &  0.060  &  0.046 &  0.012 &  0.008 &  0.005 &  0.026 &  0.019 &  0.023 &  0.030 &  0.073 &  0.003 \\
				$\sigma_{\rm mean}$	&  0.051 &  0.101 &  0.052 &  0.025  &  0.143 &  0.140 &  0.035 &  0.034 &  0.131 &  0.106 &  0.171 &  0.096 &  0.050 &  ...   \\
				$\sigma_{\rm total}$	 &  0.162 &  0.192 &  0.185 &  0.109  &  0.178 &  0.143 &  0.056 &  0.034 &  0.151 &  0.111 &  0.182 &  0.123 &  0.134 &  0.134 \\
				\hline
				\hline
				2M17542652$-$2408478 & & & & & &  & &  & & &  & & & \\
				$\sigma_{\rm T_{\rm eff}}$ &  0.066 &  0.134 &  0.114 &  0.056  &  0.212 &  0.040 &  0.044 &  0.056 &  0.036 &  0.067 &  0.057 &  0.073 &  0.080 &  ...   \\
				$\sigma_{\rm \log \textit{g}}$ &  0.105 &  0.125 &  0.034 &  0.020  &  0.016 &  0.044 &  0.042 &  0.051 &  0.002 &  0.051 &  0.016 &  0.079 &  0.072 &  ...   \\
				$\sigma_{\rm \xi_t}$	&  0.054 &  0.057 &  0.014 &  0.005  &  0.017 &  0.019 &  0.026 &  0.034 &  0.003 &  0.040 &  0.012 &  0.004 &  0.036 &  ...   \\
				$\sigma_{\rm mean}$	&  0.156 &  0.089 &  0.076 &  0.045  &  0.086 &  0.112 &  0.030 &  0.069 &  0.122 &  0.101 &  0.096 &  0.042 &  0.051 &  ...   \\
				$\sigma_{\rm total}$	 &  0.206 &  0.212 &  0.142 &  0.075  &  0.229 &  0.128 &  0.073 &  0.108 &  0.127 &  0.137 &  0.113 &  0.116 &  0.124 &  ...   \\
				\hline
				\hline
				2M17544067$-$2410464 & & & & & &  & &  & & &  & & & \\
				$\sigma_{\rm T_{\rm eff}}$ &  0.022 &  0.101 &  0.134 &  0.036  &  0.089 &  0.033 &  0.028 &  0.056 &  0.118 &  0.044 &  0.047 &  0.024 &  0.114 &  0.021 \\
				$\sigma_{\rm \log \textit{g}}$ &  0.029 &  0.057 &  0.051 &  0.006  &  0.010 &  0.035 &  0.048 &  0.056 &  0.013 &  0.057 &  0.048 &  0.013 &  0.078 &  0.071 \\
				$\sigma_{\rm \xi_t}$	 &  0.016 &  0.028 &  0.028 &  0.016  &  0.009 &  0.012 &  0.024 &  0.029 &  0.048 &  0.027 &  0.031 &  0.012 &  0.076 &  0.001 \\
				$\sigma_{\rm mean}$	 &  0.039 &  0.078 &  0.052 &  0.022  &  0.116 &  0.120 &  0.151 &  0.026 &  0.082 &  0.016 &  0.189 &  0.076 &  0.060 &  ...   \\
				$\sigma_{\rm total}$	 &  0.056 &  0.143 &  0.155 &  0.046  &  0.147 &  0.129 &  0.163 &  0.088 &  0.152 &  0.079 &  0.203 &  0.082 &  0.169 &  0.074 \\
				\hline
				\hline
				2M17543085$-$2407438 & & & & & &  & &  & & &  & & & \\
				$\sigma_{\rm T_{\rm eff}}$ &  0.025 &  0.143 &  0.119 &  0.056  &  0.073 &  0.020 &  0.057 &  0.068 &  0.103 &  0.036 &  0.013 &  0.069 &  0.025 &  0.012 \\
				$\sigma_{\rm \log \textit{g}}$ &  0.057 &  0.120 &  0.031 &  0.088  &  0.026 &  0.019 &  0.048 &  0.051 &  0.028 &  0.053 &  0.001 &  0.001 &  0.010 &  0.056 \\
				$\sigma_{\rm \xi_t}$	&  0.013 &  0.047 &  0.022 &  0.036  &  0.012 &  0.008 &  0.032 &  0.030 &  0.024 &  0.038 &  0.004 &  0.031 &  0.014 &  0.015 \\
				$\sigma_{\rm mean}$ &  0.089 &  0.121 &  0.048 &  0.031  &  0.068 &  0.103 &  0.033 &  0.074 &  0.133 &  0.099 &  0.125 &  0.001 &  0.020 &  ...   \\
				$\sigma_{\rm total}$	 &  0.109 &  0.227 &  0.134 &  0.115  &  0.104 &  0.107 &  0.088 &  0.117 &  0.172 &  0.124 &  0.126 &  0.076 &  0.036 &  0.059 \\
				\hline
				\hline
				2M17542434$-$2407356 & & & & & &  & &  & & &  & & & \\
				$\sigma_{\rm T_{\rm eff}}$ &  0.063 &  0.054 &  ...   &  0.149  &  0.120 &  0.058 &  0.081 &  ...   &  0.137 &  0.104 &  0.109 &  0.156 &  ...   &  ...  \\
				$\sigma_{\rm \log \textit{g}}$ &  0.170 &  0.117 &  ...   &  0.010  &  0.171 &  0.004 &  0.067 &  ...   &  0.024 &  0.071 &  0.036 &  0.078 &  ...   &  ...  \\
				$\sigma_{\rm \xi_t}$	&  0.165 &  0.128 &  ...   &  0.091  &  0.118 &  0.018 &  0.056 &  ...   &  0.071 &  0.095 &  0.096 &  0.007 &  ...   &  ...  \\
				$\sigma_{\rm mean}$	&  0.079 &  0.096 &  ...   &  0.011  &  0.115 &  0.156 &  0.110 &  ...   &  0.162 &  0.106 &  0.106 &  0.118 &  ...   &  ...  \\
				$\sigma_{\rm total}$	 &  0.258 &  0.205 &  ...   &  0.175  &  0.266 &  0.167 &  0.162 &  ...   &  0.225 &  0.190 &  0.183 &  0.211 &  ...   &  ...  \\				
				\hline
			\end{tabular}  \label{Table2}
		\end{center}
	\end{small}
\end{table*}

	     	\begin{table*}
	\begin{small}
		\begin{center}
			\setlength{\tabcolsep}{1.5mm}  
			\caption{Orbital elements of UKS~1.}
			\centering
			\begin{tabular}{|cccccccc|}
				\hline
				$d_{\odot} = 7.8$ kpc &  &    &   &   &  &  &  \\     			
				\hline
				$\Omega_{\rm bar}$ & $r_{peri}$ & $r_{apo}$ & $|Z_{max}|$ & $e$ & $L_{z}^{min}$ & $L_{z}^{max}$ & Orbit \\
			    	  (km s$^{-1}$ kpc$^{-1}$)& (kpc) & (kpc) & (kpc) & & ($\times10^2$ km s$^{-1}$ kpc) & ($\times10^2$ km s$^{-1}$ kpc)& \\
				\hline     	
				\hline		
				31 &  0.01$\pm$0.01 &   1.40$\pm$0.25 &  0.29$\pm$0.06 &  0.97$\pm$0.02 &  $-$12.0$\pm$9.0  &    9.0$\pm$ 6.0 & P-R\\
				41 &  0.01$\pm$0.01 &   1.42$\pm$0.23 &  0.49$\pm$0.18 &  0.97$\pm$0.02 &  $-$13.0$\pm$6.5  &    7.0$\pm$ 7.0 & P-R\\
				51 &  0.01$\pm$0.06 &   1.42$\pm$0.23 &  0.28$\pm$0.05 &  0.97$\pm$0.07 &  $-$11.0$\pm$9.5  &    5.0$\pm$ 7.0 & P-R\\
				\hline
				\hline
				$d_{\odot} = 15.9$ kpc &  &    &   &   &  &   &  \\
				\hline
				$\Omega_{\rm bar}$ & $r_{peri}$ & $r_{apo}$ & $|Z_{max}|$ & $e$ & $L_{z}^{min}$ & $L_{z}^{max}$ & Orbit \\
			    	  (km s$^{-1}$ kpc$^{-1}$)& (kpc) & (kpc) & (kpc) & & ($\times10^2$ km s$^{-1}$ kpc) & ($\times10^2$ km s$^{-1}$ kpc)& \\
				\hline
				\hline
				31 &  0.17$\pm$0.48 &   8.88$\pm$1.80 &  5.31$\pm$2.49 &  0.95$\pm$0.08 &  $-$21.0$\pm$24.0 &    3.0$\pm$30.0 & P-R\\
				41 &  0.24$\pm$0.51 &   9.29$\pm$1.55 &  5.36$\pm$2.61 &  0.94$\pm$0.08 &  $-$28.0$\pm$21.5 & $-$6.0$\pm$30.5 & Prograde \\
				51 &  0.25$\pm$0.44 &   9.22$\pm$1.45 &  6.00$\pm$2.50 &  0.94$\pm$0.07 &  $-$22.0$\pm$20.0 & $-$6.0$\pm$20.5 & Prograde \\
				\hline
			\end{tabular}  \label{Table3}
		\end{center}
	\end{small}
\end{table*}	 	

\clearpage

\begin{appendix}

\section{The absolute PMs of UKS~1}	
\label{section8}

For the orbit computations, we adopt the proper motions measured by VIRAC \citep{Smith2018}, which were examined in detail to find a good size sample of UKS~1 stars with the least possible contamination from field stars, so that a robust determination of the cluster's proper motion can be made. The following cuts were performed, yielding a sample of 2297 stars:

\begin{itemize}
	\item[$-$] $11.5<K_s<15.8$ and  $J-Ks>2$: The brightest cut in magnitude rejects stars affected by saturation effects while the faintest cut avoids excedingly large errors in proper motions and significant contamination from field stars. The cut in colour corresponds to the known locus of UKS~1 stars.
	
	\item[$-$] Distance to cluster centre $r<1.35'$: An examination of $\mu_\delta$ versus $r$ for a bright red sample ($11.5<K_s<12.8$ and $J-Ks>2$) revealed that a cut in radius is necessary to avoid significant contamination from the bulge red giants, at the cost of losing the outskirts of the cluster. In the innermost radius there will be some contamination, but it is minimized by the large number of UKS~1 stars.
	
	\item[$-$] Absolute value of normalized proper motion in declination less than 5: To properly account for the effects of proper motion errors, we estimate, for the sample selected by the two previous items, an approximate mean proper motion in declination for the cluster of -2.2 mas yr$^{-1}$, and we use that to compute: 
	$\displaystyle \mu_{\delta,\mbox{norm}}=\frac{\mu_\delta-(-2.2)}{\epsilon_{\mu_\delta}}$. This quantity effectively brings closer to zero many members of UKS~1, despite their span in proper motion errors. This allows us to build a large and robust sample dominated by UKS~1 stars that properly reflects the distribution of VIRAC proper motion errors. This value also spreads a substantial number of field outliers to large values far from zero. This procedure was only done in $\mu_\delta$ because field and cluster populations only separate far enough in this coordinate for it to work successfully.
	
	\item[$-$] $|\mu_\alpha\cos(\delta)|<30$ mas yr$^{-1}$: Once all previous steps were applied, we finally cut evident outliers in $\mu_\alpha\cos(\delta)$, so that the data spans around the observed mean value to an extent similar to $\mu_\delta$.
	
\end{itemize}

The proper motions of the final resulting sample of 2297 stars were analysed using a quantile-quantile Q-Q plot, to compare their distribution to a normal standard N(0,1) distribution, in order to see if it properly describes the data and to compute its parameters (mean, standard deviation, and its corresponding errors). In a Q-Q plot, data that is normally distributed will end up following a straight line, whose zero point is the mean and the slope is the standard deviation of the data. We found that the UKS~1 selected sample has heavier tails than a normal distribution, not unexpectedly. On the other hand, the innermost $\sim$80\% behaves well, and fitting a straight line to it yielded the following results for the mean proper motion of UKS~1: $(\mu_\alpha\cos(\delta),\mu_\delta)=(-2.77,-2.43) \pm (0.23, 0.16)$ mas yr$^{-1}$  . The errors quoted here were computed by dividing the obtained standard deviations $(2.60, 2.29)$  mas yr$^{-1}$ by the square root of the number of data points used in the fit. The formal errors of this procedure for the mean value are in fact much smaller, 0.005 mas yr$^{-1}$ per year, but we believe this value underestimates the real quality of the data, and prefer to adopt the chosen error estimate as more representative of the data. The $K_s$ versus $J-K_s$ CMD of the above selected sample (2297 stars) looks as expected for UKS~1, and confirms our selection to be reasonable.

Our proper motion value is similar but slightly off in $\mu_\delta$ from the one obtained by \citet{Baumgardt2019} using GAIA DR2: $(-2.59,-3.42)\pm(0.52,0.44)$. Our analysis indicates that contamination from bulge red giants biases $\mu_\delta$ to lower values, which actually forced us to limit the sample to the innermost portion of the cluster in an effort to minimize it. This strategy was not a factor for \textit{GAIA} DR2 because its data are much shallower than those of VIRAC. This also confirms, the enormous value (because of their depth) and exquisite quality (because of it being consistent with GAIA DR2) of the VIRAC proper motions.		

 \section{Differential reddening correction}
 \label{section9}
 
    The differential reddening correction was performed using giant stars, and by adopting the reddening law of \citet{Cardelli1989} and \citet{Donnell1994} and a total-to-selective absorption ratio $R_{V}=3.1$. For this purpose, we selected all RGB stars within a radius of 5' from the cluster centre and that have proper motions compatible with that of UK1 within 1 mas yr$^{-1}$. First, we draw a ridge line along the RGB, and for each of the selected RGB stars we calculated its distance from this line along the reddening vector. The vertical projection of this distance gives the differencial ${\rm A_{K}}$ absorption at the position of the star, while the horizontal projection gives the differential ${\rm E(J-K)}$ reddening at the position of the star. After this first step, for each star of the field we selected the three nearest RGB stars, calculated the mean differential ${\rm A_{K}}$ absorption and the mean differential ${\rm E(J-K)}$ reddening, and finally subtracted these mean values from its ${\rm J-K}_{\rm s,VVV}$ colour and ${\rm K_{s, VVV}}$ magnitude. We underline the fact that the number of reference stars used for the reddening correction  is a compromise between having a correction affected as little as possible by photometric random error and the highest possible spatial resolution. The differential reddening corrections for VVV bands is listed in Table \ref{Table2}.  
	
\end{appendix} 

\end{document}